\documentclass[superscriptaddress,amsmath,amssymb,amsfonts,onecolumn]{revtex4}
\usepackage{soul}
\usepackage[percent]{overpic}
\usepackage{amsmath}
\usepackage{subfigure}
\usepackage{graphicx}% Include figure files
\usepackage{dcolumn}% Align table columns on decimal point
\usepackage{bm}% bold math
\usepackage{amssymb}
\usepackage{physics}
\usepackage{soul}
\usepackage{xcolor}
\usepackage{xparse}
\usepackage{color}
\usepackage{enumitem}
\usepackage{enumitem}
\usepackage{pgfplots}
\pgfplotsset{compat=1.9}
\usepackage{tikz}
\usepackage{hyperref}
\usepackage{float}
\usepackage{orcidlink}
\newcommand{\R}{\mathbb{R}}

\usepackage[normalem]{ulem}

\begin{document}

%\preprint{APS/123-QED}

\title{Homoclinic and heteroclinic solutions of the nonlinear
  Schr\"odinger equation with a localized Wadati potential}

\author{Sathyanarayanan Chandramouli \orcidlink{0000-0002-9520-1870}} \email{sathyanaraya@umass.edu}
\affiliation{Department of Mathematics and Statistics, University of
  Massachusetts at Amherst}

\author{Patrick Sprenger \orcidlink{0000-0001-6654-8974}}
\email{sprenger@ucmerced.edu} \affiliation{Department of Applied
  Mathematics, University of California at Merced}

\author{Mark A.~Hoefer \orcidlink{0000-0001-5883-6562}}
\email{hoefer@colorado.edu} \affiliation{Department of Applied
  Mathematics, University of Colorado at Boulder}

\date{\today}

\begin{abstract}
  Stationary solutions asymptoting to nonlinear plane waves of the
  nonlinear Schr\"odinger equation with a localized, PT-symmetric,
  complex linear potential are characterized.  The potential includes
  both a spatially varying gain-loss profile and a repulsive real
  part, generated by a Wadati potential function, that support the
  existence of homoclinic and heteroclinic solutions that asymptote to
  the same or different, respectively, nonlinear plane waves in the
  far field.  Asymptotic analysis and numerical simulations are used
  to examine solution existence, bifurcations, and structure.  Such
  solutions play an important role in resonant nonlinear wave
  generation of dispersive media with localized gain and loss.
  \end{abstract}

\maketitle
%\tableofcontents
\section{Introduction}
\label{sec:introduction}
Nonlinear wave phenomena in non-Hermitian systems have been the
subject of intense research activity in recent decades since the
seminal paper in quantum mechanics exhibiting Hamiltonian-like
properties for a non-Hamiltonian Schr\"odinger equation
\cite{bender1998real}.  The class of PT-symmetric linear operators
with gain and loss that exhibit only real spectra has since been
expanded upon
\cite{el2018non,konotop2016nonlinear,zyablovsky2014pt,bender2019pt,hang2013pt,ruter2010observation}.
These systems have been realized in other areas of physics beyond
quantum mechanics including optics
\cite{chong2010coherent,ramezani2010unidirectional,makris2017wave},
metamaterials \cite{smith2004metamaterials}, non-Hermitian acoustics
\cite{gu2021controlling,achilleos2017non,zhang2021acoustic,rivet2018constant}
and exciton-polariton condensates
\cite{lien2015multistability,byrnes2014exciton}. For a review of some
of this work, see \cite{konotop2016nonlinear}.  A model of
non-Hermitian nonlinear waves is the variable coefficient nonlinear
Schr\"odinger (NLS) equation
\begin{equation}
  \label{eq:1}
  \begin{split}
    i \psi_t &= - \frac{1}{2} \psi_{xx} + V(x) \psi + |\psi|^2 \psi,
    \\ 
             &\psi:~\mathbb{R} \times (0,\infty) \to \mathbb{C},
  \end{split}
\end{equation}
where
\begin{equation}
  \label{eq:58}
  V(x) = \frac{1}{2} \left ( -w(x)^2 + i w'(x) \right )
\end{equation}
is the complex linear potential for $w(x)$ a real-valued, smooth
Wadati potential function. This potential was first introduced by
Wadati \cite{wadati2008construction} in the construction of
PT-symmetric potentials for the Schr\"odinger operator using inverse
scattering theory for the modified Korteweg-de Vries
equation. Equation \eqref{eq:1} with potential \eqref{eq:58} is a
nonlinear generalization of non-Hermitian, Schr\"odinger-type dynamics
that has generated significant interest
\cite{cole2016modulational,komis2020equal,makris2019two,makris2020scattering,makris2017wave,makris2015constant,makris2016constant,ossi2022topological,rivet2018constant,zezyulin2022nonlinear,chandramouli2023dispersive,nixon2016nonlinear}.

By virtue of it's sign, the real part of $V(x)$ corresponds
to a repulsive potential.  The imaginary part corresponds to gain in regions of space where $w(x)$ is increasing and loss where $w(x)$ is decreasing.  In this work, we assume the
following additional properties:
\begin{enumerate}
\label{Properties}
\item[\{1\}] PT-symmetry: $w(-x) = w(x)$
  \cite{wadati2008construction} this symmetry implies invariance under
  space-time reversal and complex conjugation. Thus, if $\psi(x,t)$ is a
  solution to Equation \eqref{eq:1}, so is $\psi^{*}(-x,-t)$.
\item[\{2\}] decay: $|w(x)| < C e^{- \alpha |x|}$ for some
  $\alpha, C > 0$.
\item[\{3\}] smooth potential: $w \in C^1(\mathbb{R})$, $w'(x) = 0$ only
  when $x = 0$.
\item[\{4\}] normalization: $w(0) = 1$ .
\end{enumerate}
We refer to $w(x)$ as the \textit{Wadati potential function} and
$V(x)$ in \eqref{eq:58} as the \textit{complex linear potential}.  Due
to a symmetry of \eqref{eq:1}, introduced later, the normalization
condition \{4\} can be relaxed by a rescaling of the independent and
dependent variables. The positive pulse shape resulting from
properties \{3,4\} implies that gain is only active in the region
$x < 0$ and loss is active for $x > 0$.  A family of Wadati potential
functions that we will closely examine in this work is
\begin{equation}
  \label{eq:wadati_sech_potential}
  w(x) = \mathrm{sech}^n(\epsilon x), \quad n \ge 1, \quad \epsilon > 0 ,
\end{equation}
whose exponential decay rate \{2\} is $\alpha = n \epsilon$ and its
full width at half maximum
$\Delta = \mathrm{cosh}^{-1}\left (4^{1/n}\right )/\epsilon$ is
inversely proportional to $\epsilon$.

When $w(x) > 0$, the linear Schr\"odinger operator
$-\mathrm{d}^2/\mathrm{d}x^2+V(x)$ with complex linear potential
\eqref{eq:58} possesses an all-real eigenspectrum
\cite{barashenkov2016exactly,tsoy2014stable,klaus2002purely}. The NLS
equation \eqref{eq:1} with complex linear potential \eqref{eq:58}
exhibits continuous families of stationary nonlinear solutions and
constant intensity plane wave solutions with eigenvalue quartets in
the linearized spectrum
\cite{makris2015constant,zezyulin2022nonlinear} that are reminiscent
of solutions to the conservative, Hamiltonian NLS equation with a
real-valued linear potential $V(x)$.  If $w(x)$ is not an even
function, PT-symmetry is broken so the reality of the associated
Schr\"odinger operator's spectrum is not guaranteed. However, even in
the absence of PT-symmetry, properties reminiscent of conservative,
Hamiltonian systems have been observed in these gain-loss systems. For
example, continuous families of nonlinear stationary solutions are
known to exist
\cite{nixon2016nonlinear,barashenkov2016exactly,zezyulin2022nonlinear}. Two-dimensional
extensions of eq.~\eqref{eq:1} with complex linear potential
\eqref{eq:58} have also been studied
\cite{makris2015constant,yang2021analytical}.

When $V(x)$ is real-valued, eq.~\eqref{eq:1} is known as the
Gross-Pitaevskii (GP) equation that models light propagation in
nonlinear media with normal dispersion \cite{boyd_nonlinear_2003} and
Bose-Einstein condensates (BECs) subject to a repulsive
self-interaction \cite{pitaevskii_bose-einstein_2003}.  By a suitable
change of variables, the GP equation admits a hydrodynamic
interpretation in which nonlinear plane wave solutions
$\psi(x,t) = \sqrt{\rho_0}e^{i(u_0x-(\rho_0+u_0^2/2) t)}$ for
$V(x) \equiv 0$ are analogous to a compressible fluid with constant
density $\rho_0 > 0$ and velocity $u_0 \in \mathbb{R}$.  Gradient
catastrophe is regularized by dispersion leading to the formation of
dispersive shock waves (DSWs) that are oscillatory, expanding
nonlinear wavetrains, a prominent multiscale feature of dispersive
hydrodynamics \cite{el_dispersive_2016-1}.

The existence of constant intensity plane wave solutions to
eq.~\eqref{eq:1} with complex linear potential \eqref{eq:58}
\cite{makris2015constant}
\begin{equation}
  \label{eq:15}
  \psi(x,t) = \sqrt{\rho_0} e^{i (\int^x
    w(x')\,\mathrm{d}x' - \rho_0 t)}
\end{equation}
opened the door to the study of non-Hermitian dispersive hydrodynamics
in \cite{chandramouli2023dispersive} where the resonant generation of
nonlinear waves due to the presence of gain and loss was numerically
observed. Such dynamics are reminiscent of transcritical flow of a
fluid past an inhomogeneity where sub and supercritical velocities
exhibit steady, homoclinic flow profiles whereas intermediate,
transcritical velocities lead to the spontaneous generation of DSWs
and solitary wavetrains \cite{grimshaw_resonant_1986-1}.  Resonant
nonlinear wave generation occurs because the transcritical region is
marked by the non-existence of steady homoclinic flow
patterns. Instead, steady heteroclinic flow patterns emerge, which do
not satisfy the far-field boundary conditions.  This phenomenon has
been extensively studied in various conservative settings arising in
fluid mechanics, optics, and superfluids
\cite{grimshaw_resonant_1986-1,leszczyszyn_transcritical_2009,lee_upstreamadvancing_1990,lee1989experiments,hakim_nonlinear_1997-1,engels_stationary_2007,wu1981long}. In
such contexts, when the background “flow” speed in the medium
approaches the corresponding long-wave speed, it can lead to the
emergence of nonlinear excitations such as DSWs and solitary
wavetrains. Specifically in superfluid systems, this resonant
generation of nonlinear waves has been observed to occur over a range
of flow speeds centered around the long-wave speed—commonly referred
to as the ``transcritical" interval—ultimately resulting in the
breakdown of superfluidity \cite{engels_stationary_2007}. The
breakdown of superfluidity due to an underlying resonance has been a
subject of fundamental importance, with a long history dating back to
early investigations in He-II (cf. \cite{gorter1961progress,
  wilks1967properties})

Inspired by the problem of resonant nonlinear wave generation in
non-Hermitian media and the possibility of dispersive hydrodynamics
subject to gain and loss, we seek to identify steady flow patterns in
the form of homoclinic and heteroclinic solutions of eq.~\eqref{eq:1},
with the localized complex linear potential \eqref{eq:58} subject to
plane wave boundary conditions. This investigation into the existence
of stationary solutions provides a foundation for the study of the
non-Hermitian transcritical flow problem. Asymptotic and numerical
analysis are used to assess the existence, bifurcation, and properties
of these solutions.  It is demonstrated that this PT-symmetric,
gain-loss nonlinear wave system supports a variety of stationary
solutions that do not exist in the absence of gain and loss.

The manuscript is structured as follows.  Hydrodynamic variables are
introduced in section \ref{sec:disp-hydro-formulation}, followed by the formulation of steady
solutions in section \ref{sec:steady-solut-gener}.  In section
\ref{sec:homoclinic-solutions-1}, we characterize homoclinic solutions
to the same plane wave boundary conditions whereas in section
\ref{sec:heter-solut}, we study heteroclinic solutions subject to
different plane wave boundary conditions.  Section
\ref{sec:discussion} presents some discussion and concluding points.
The appendices in section \ref{Hydraulic-homoclinic-repulsive} present a detailed discussion of the hydraulic limit for the Wadati case, followed by
a brief analysis of slowly varying, steady solutions to
eq.~\eqref{eq:1} with the real-valued, repulsive linear potential
$V(x) = -w^2(x)$ to contrast with the complex-valued linear potential
\eqref{eq:58}.

\section{Dispersive hydrodynamic formulation}
\label{sec:disp-hydro-formulation}

Introducing the transformation $\psi = \sqrt{\rho} e^{i \phi}$ into
\eqref{eq:1} and equating real and imaginary parts results in the
hydrodynamic formulation of the NLS equation with Wadati potential
function $w(x)$
\begin{subequations}
  \label{eq:hydro_formulation}
  \begin{align}
    \label{eq-rho-ev}
    \rho_t + (\rho \phi_x)_x
    &= \rho w',\\
    \label{eq-phi}
    \phi_t + \frac{1}{2} |\phi_x|^2 + \rho
    &= \frac{1}{4} \left [ \frac{\rho_{xx}}{\rho} - \frac{\rho_x^2}{2
      \rho^2} \right ] + \frac{1}{2} w^2 ,
  \end{align}
  where $' \equiv \frac{\mathrm{d}}{\mathrm{d}x}$ denotes a total
  spatial derivative.  Equation \eqref{eq-rho-ev} for the evolution of
  the hydrodynamic density $\rho$ is a continuity equation subject to
  the source/sink or gain/loss term $\rho w'$ due to the imaginary
  part of the complex linear potential \eqref{eq:58}.  Equation
  \eqref{eq-phi} is a dispersive analog of the Bernoulli equation for
  a compressible fluid.  Setting $u = \phi_x$ and taking $\partial_x$
  of eq.~\eqref{eq-phi} results in
\begin{equation}
  \label{u-eqn}
  u_t+ \left ( \frac12 u^2 +\rho 
  \right )_x = \frac{1}{4} \left ( \frac{\rho_{xx}}{\rho} -
    \frac{\rho_x^2}{2 \rho^2} \right )_x + \frac12 \left ( w^2 \right
  )',
\end{equation}
a dispersively regularized equation for the hydrodynamic velocity $u$
subject to the conservative, repulsive force $(w^2)'/2$ due to the
real part of the complex linear potential \eqref{eq:58}.  We can
combine \eqref{eq-rho-ev} and \eqref{u-eqn} to obtain the conservation
law, first identified in \cite{nixon2016nonlinear}
\begin{equation}
  \label{m-eqn}
  m_t + \left ( um + \frac{1}{2} \rho^2 \right )_x
  = \frac{1}{4} \left [ \rho (\log
    \rho)_{xx} \right ]_x ,
\end{equation}
\end{subequations}
for the shifted hydrodynamic momentum
\begin{equation}
  \label{eq:41}
  m = \rho(u-w) .
\end{equation}
Equations \eqref{eq-rho-ev}, \eqref{u-eqn}, and \eqref{m-eqn}
constitute the hydrodynamic equations for the density $\rho$, the
velocity $u$, and the shifted momentum $m$ of the nonlinear
Schr\"odinger fluid.  Thus, the hydrodynamic interpretation
\eqref{eq:hydro_formulation} of the NLS equation~\eqref{eq:1} subject
to the complex linear potential \eqref{eq:58} results in a dispersive
regularization of compressible, Eulerian fluid dynamics subject to a
conservative, repulsive force and gain/loss of the fluid density.

Equations \eqref{eq-rho-ev}, \eqref{u-eqn}, and \eqref{m-eqn} are
invariant under the transformations
\begin{equation}
  \label{symmetries}
  \begin{split}
    \rho \to B \rho, \quad u \to B^{1/2} u,
    &\quad w \to B^{1/2} w, \quad m \to B^{3/2} m, \quad  x \to
     B^{-1/2} x, \quad t \to B^{-1}t , \quad B > 0 , \\
    &u \to -u, \quad w \to -w, \quad m \to -m, \quad x \to -x . 
  \end{split}
\end{equation}
By virtue of these symmetries, we will, without loss of generality,
assume the aforementioned normalization $w(0) = 1$ unless otherwise
stated.

\section{Steady solutions}
\label{sec:steady-solut-gener}

Our interest is in steady solutions of eq.~\eqref{eq:hydro_formulation} in which
$\phi_t = \mu \in \R$ in eq.~\eqref{eq-phi} and $\rho_t = u_t = m_t = 0$ in \eqref{eq-rho-ev},
\eqref{u-eqn}, and \eqref{m-eqn}.  Then, we can express
eq.~\eqref{eq-rho-ev} as
\begin{subequations}
  \label{eq:49}
  \begin{equation}
    \label{eq:4}
    m' = - \rho'w .
  \end{equation}
  Equations \eqref{eq-phi} and \eqref{m-eqn} become
  \begin{align}
    \label{eq:56}
    2\mu + u^2 - w^2 + 2\rho
    &= \frac{\rho''}{2\rho} - \frac{(\rho')^2}{4 \rho^2},
    \\
    \label{eq:57}
    \left ( um +
    \frac{1}{2} \rho^2 \right )'
    &= \frac{1}{4} \left [ \rho'' - \frac{(\rho')^2}{\rho} \right ]' .
  \end{align}
\end{subequations}

Two classes of solutions of \eqref{eq:49} are of interest
distinguished by their far-field boundary conditions
\begin{subequations}
  \label{eq:43}
  \begin{align}
    \label{eq:47}
    &\mathrm{homoclinic:} \quad \rho \to \rho_0, \quad u \to
      u_0 \ge 0, \quad |x| \to \infty, \\
    \label{eq:3}
    &\mathrm{heteroclinic:} \quad \rho \to \rho_\pm, \quad u
      \to u_\pm, \quad x  \to \pm \infty,
  \end{align}
\end{subequations}
where, additionally, $\rho'(x) \to 0$ as $x \to \infty$ or
$x \to -\infty$.  These boundary conditions correspond to plane waves
in the far field for the complex wavefunction $\psi$ in
eq.~\eqref{eq:1}.  The analysis of homoclinic solutions is restricted
to non-negative far-field velocities $u_0 \ge 0$ because the case of
negative velocities gives rise to different regimes that warrant a
separate study.

Integrating eq.~\eqref{eq:57} with integration constant $A$, dividing
by $\rho$, subtracting it from eq.~\eqref{eq:56}, rearranging, and
combining with \eqref{eq:4} results in the third order system of
ordinary differential equations (ODEs)
\begin{equation}
  \label{eq:29}
  \rho'' = 2 \dot F(\rho) + 4 w m, \quad m' = - w \rho' ,
\end{equation}
where the overdot denotes differentiation with respect to $\rho$
($\dot F \equiv {\rm d} F/{\rm d} \rho$) and
\begin{equation}
  \label{eq:63}
  F(\rho) = \rho^3 + 2\mu \rho^2 + 2 A \rho ,
\end{equation}
is a cubic polynomial with three roots
\begin{equation}
  \label{eq:66}
  \rho \in \left \{ 0, -\mu \pm \sqrt{\mu^2 - 2A} \right \} .
\end{equation}
For the class of potentials considered in this paper with even-parity,
the stationary ODE system in eq.~\eqref{eq:29} is invariant to the
reflection $x \rightarrow -x$. This underlying symmetry allows for
heteroclinic solutions of either polarity, with $\rho'(0) > 0$ or $\rho'(0) < 0$. 

Solutions of the nonlinear boundary value problem (BVP) consisting of
eq.~\eqref{eq:29} subject to one of the boundary conditions
\eqref{eq:43} are the focus of this study.

\begin{figure}
    \centering
    \includegraphics[width=\linewidth]{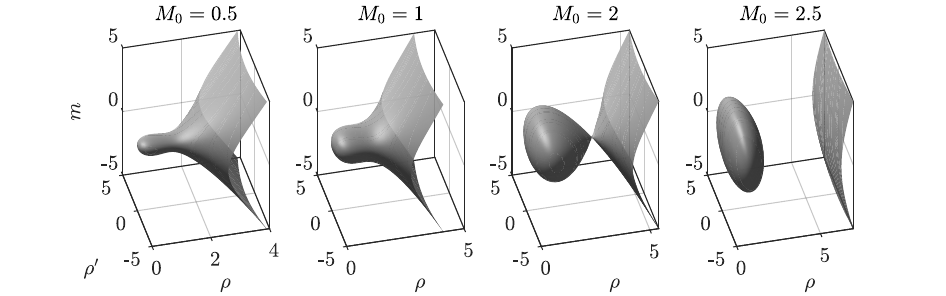}
    \caption{Zero energy surfaces for different Mach numbers $M_0$. }
    \label{fig:zero_energy_surface}
\end{figure}

The parameters $\mu$ and $A$ are real constants determined by the
boundary conditions \eqref{eq:43}.  Multiplying eq.~\eqref{eq:56} by
$2\rho$, subtracting the first of eq.~\eqref{eq:29}, and rearranging
results in the zero energy integral, satisfied by all smooth solutions
to the BVP \cite{konotop2014families}
\begin{equation}
  \label{eq:62}
  E = \frac{1}{4} \rho'(x)^2 + m(x)^2 - F(\rho(x)) = 0 , \quad
  x \in \R .
\end{equation}
The integral \eqref{eq:62} does not explicitly depend on $w(x)$ so is
parametrized by the constants $\mu$ and $A$.  We can characterize this
energy surface in the $\rho$-$\rho'$-$m$ space in terms of one
constant parameter by applying one of the boundary conditions in
\eqref{eq:43} and dividing the density by its corresponding boundary
value $\rho_0$, say as $x \to \infty$.  Note that we are temporarily
relaxing the unit Wadati potential function assumption to
$w(0) = \rho_0^{-1/2}$ in favor of a unit density boundary condition
but can always recover unit potential by use of the symmetry
\eqref{symmetries}. With this normalization, the geometry of the
surface \eqref{eq:62} depends on the ratio of the fluid speed to the
long wave speed or Mach number
\begin{equation}
   \label{eq:Mach_number}
   M_0 = \frac{|u_0|}{\sqrt{\rho_0}},
\end{equation}
where $u_0$ and $\rho_0$ are constants from \eqref{eq:43} as
$x \to \infty$ and $\mu = -(M_0^2/2+1)$, $A = (M_0^2 + 1/2)$.  The
far-field flow is \textit{subsonic}, \textit{sonic}, or
\textit{supersonic} when $M_0 < 1$, $M_0 = 1$, or $M_0 > 1$,
respectively. The energy surface \eqref{eq:62} for different Mach
numbers $M_0$ is shown in Fig.~\ref{fig:zero_energy_surface}.  At the
coarsest level, for $0 < M_0 \le 2$, the surface is connected while
for $M_0 > 2$, the surface is disconnected.  In future sections, we
will present some trajectories of steady solutions on the zero energy
surface \eqref{eq:62}.

The ODE system \eqref{eq:29} admits a variational formulation in which
eq.~\eqref{eq:29} is the Euler-Lagrange equation of
\begin{subequations}
  \label{eq:variational}
  \begin{equation}
    \label{eq:64}
    J = \int_{\mathbb{R}} \left ( \frac{1}{4} (\rho')^2 + F(\rho) -
      m(\rho)^2 \right )\,\mathrm{d}x, 
  \end{equation}
  subject to the constraint \eqref{eq:62} and
  \begin{equation}
    \label{eq:65}
    \dot{m}(\rho) = -w.
  \end{equation}
\end{subequations}

\subsection{Hydraulic regime}
\label{sec:hydraulic-regime}

When $w$ is slowly varying in space, we introduce its spatial scaling
according to
\begin{equation}
  \label{eq:32}
  w = w(X), \quad X = \epsilon x, \quad
  |w'(X)/w(X)| = \mathcal{O}(1), \quad 0 < \epsilon \ll 1. 
\end{equation}
Assuming that the hydrodynamic variables are also slowly varying
$\rho = \rho(X)$, $u = u(X)$, $m = m(X)$, this scaling leads to the
singularly perturbed ODEs \eqref{eq:29}
\begin{equation}
  \label{eq:24}
  \epsilon^2 \rho'' = 2 \dot F(\rho) +
  4 w m, \quad m' = - w \rho' ,
\end{equation}
where, now $' \equiv \mathrm{d}^2/\mathrm{d}X^2$.  Setting
$\epsilon = 0$ corresponds to the dispersionless limit of the NLS
equation \eqref{eq:1} with complex linear potential \eqref{eq:58}.

Setting $\epsilon = 0$ in eq.~\eqref{eq:24}, the system reduces to the
algebraic equations
\begin{align}
  \label{eq:9}
    F(\rho) - m^2 = 0,\\
    \label{eq:9b}
    \quad \dot F(\rho) + 2 w m = 0 ,
\end{align}
which simplify to the quartic polynomial in $\rho$ 
\begin{equation}
  \label{eq:22}
  4 w^2 F(\rho) - \dot F(\rho)^2 = 0 ,
\end{equation}
that determines the dependence of $\rho$ on $w$ and, through
\eqref{eq:9b}, the dependence of $m$ on $w$.  In the hydrodynamic
context, this is referred to as the \textit{hydraulic approximation}
\cite{grimshaw_resonant_1986-1,hakim_nonlinear_1997,leszczyszyn_transcritical_2009}
and we will adopt this terminology henceforth.
% Steady solutions in the hydraulic regime are discussed in detail in Sec. \ref{sec:hydraulic-solution}

% In the next section, homoclinic solutions satisfying \eqref{eq:29} and
% \eqref{eq:47} are analyzed, followed by heteroclinic solutions
% satisfying \eqref{eq:29} and \eqref{eq:3} in section
% \ref{sec:heter-solut}.

%%%%%%%%%%%%%%%%%%%%%%%%%%%%%%%%%%%%%%%%%%%%

\section{Homoclinic solutions}
\label{sec:homoclinic-solutions-1}

Applying the boundary conditions \eqref{eq:47} and decay of
derivatives to eqs.~\eqref{eq:56} and \eqref{eq:57} results in
\begin{equation}
    \label{eq:7}
    \mu = -\frac{1}{2} u_0^2 - \rho_0 < 0 , \quad A = \rho_0 u_0^2 +
    \frac{1}{2}\rho_0^2 > 0 .
\end{equation}
Equation \eqref{eq:29} admits the constant solution
\begin{equation}
  \label{eq:12}
  \rho(x) = \rho_0, \quad m(x) = 0, \quad \mathrm{when} \quad u_0 = 0
  , 
\end{equation}
corresponding to the constant intensity plane wave \eqref{eq:15}
\cite{makris2015constant,makris2016constant,rivet2018constant,cole2016modulational,makris2019two}.
The aim of this section is to generalize this plane wave solution to
positive far-field hydrodynamic velocity $u_0 > 0$.

Even homoclinic solutions exhibiting an extremum in density at the maximum
of $w(x)$ must satisfy $\rho'(0) = 0$ so that $m'(0) = 0$ also and,
according to \eqref{eq:41},
\begin{align}
  \label{eq:38}
  &0 = m'(0) = \rho'(0)(u(0)-1) + \rho(0)(u'(0)-w'(0)) = \rho(0) u'(0) ,
\end{align}
which implies $u'(0) = 0$ also. Such homoclinic solutions expressed in
the complex form of eq.~\eqref{eq:1}
\begin{equation}
  \label{eq:8}
  \psi(x,t) = \sqrt{\rho(x)}\exp\left [ i\int_0^{x} u(y) dy \right
  ] e^{i\mu t},
\end{equation}
satisfy PT-symmetry: $\psi^{*}(x,t)=\psi(-x,-t)$.  The zero energy
integral \eqref{eq:62} implies
\begin{equation}
  \label{eq:21}
  F(\rho(0)) - m(0)^2 = 0 ,
\end{equation}
where the sign of 
\begin{equation}
  \label{eq:23}
  \rho''(0) = 2 \dot F(\rho(0)) + 4 m(0)
\end{equation}
determines whether the extremum is a maximum or minimum.  Using
$w'(0) = 0$, $w(0) = 1$, and eq.~\eqref{eq:29},
\begin{equation}
  \label{eq:34}
  m''(0) = - \rho''(0),
\end{equation}
so that if $\rho''(0) > 0$, a minimum in density corresponds to a
maximum in momentum.  We will refer to such solutions as
\textit{depression waves}.  Similarly, if $\rho''(0) < 0$, a maximum
in density corresponds to a minimum in momentum.  Such a solution is
called an \textit{elevation wave}.  The solution \textit{polarity} is
then either negative for a depression wave or positive for an
elevation wave.
% We will exclusively study even depression and elevation wave
% solutions.  \MH{correct?}

% In the remainder of this section, we first gain some analytical
% understanding of homoclinic solutions by analysis of the far-field
% \ref{sec:linearized-far-field}, the hydraulic regime
% \ref{sec:hydraulic-solution}, the large \ref{sec:large-density-regime}
% and small \ref{sec:weakly-nonl-solut} density limits, bifurcations
% from constant intensity waves \ref{sec:depr-homocl-orbits}, and the
% exponential asymptotics of resonant elevation waves
% \ref{sec:elev-homocl-orbits}. Then, we perform extensive numerical
% studies in sec.~\ref{sec:numerical-solutions} to compare with and
% complement the analytical findings.

\subsection{Linearized, far-field analysis}
\label{sec:linearized-far-field}

Assuming symmetric, monotonic decay from $x = 0$ of both $\rho(x)$ and
$m(x)$, we investigate the far-field behavior of homoclinic solutions
by letting $\rho(x) = \rho_0 + R(x)$ and $m(x) = \rho_0 u_0 + M(x)$
and, further, assuming that $w(x) = B e^{-\alpha x}$ as
$x \to \infty$.  Then $R$ and $M$ satisfy the linearized,
inhomogeneous equations in the far-field
\begin{equation}
  \label{eq:6}
  R'' - \gamma(\rho_0) R = 4 B \rho_0 u_0 e^{-\alpha x}, \quad M'
  = - B e^{-\alpha x} R', \quad x \to \infty,
\end{equation}
with the general solution as $x \to \infty$
\begin{subequations}
  \label{eq:16}
  \begin{align}
    \label{eq:19}
    R(x) &= \frac{4 B \rho_0 u_0}{\alpha^2 - \gamma}
           e^{-\alpha x} + R_- e^{- \sqrt{\gamma} x} + R_+
           e^{\sqrt{\gamma} x}, \\
    \label{eq:11}
    M(x) &= - \frac{2 B^2 \rho_0 u_0}{\alpha^2 - \gamma} e^{-2
           \alpha x} + B\sqrt{\gamma} e^{-\alpha x} \left ( -
           \frac{R_-}{\alpha + \sqrt{\gamma}} e^{-\sqrt{\gamma}x} +
           \frac{R_+}{\alpha - \sqrt{\gamma}} e^{\sqrt{\gamma} x}
           \right ), 
  \end{align}
\end{subequations}
where
\begin{equation}
  \label{eq:20}
  \gamma = 2 \ddot{F}(\rho_0) = 4(\rho_0 - u_0^2) 
\end{equation}
and $R_\pm$ are real numbers if $\gamma \ge 0$ and complex conjugates
if $\gamma < 0$. If we assume that $R_\pm = 0$ and the far-field is
supersonic or sonic ($u_0 \ge \sqrt{\rho_0}$), then
$\rho(x) = \rho_0 + R(x) > \rho_0$ as $x \to \infty$ so that any
homoclinic solution for these boundary conditions is an elevation
wave.  We shall see that supersonic boundary conditions coincide with
a short wave resonance in density so that $\gamma < 0$ and
$R_\pm \ne 0$.

When the far-field is subsonic so that $\gamma > 0$, then $R_+ = 0$
for decaying solutions.  Further assuming that
$\alpha < \sqrt{\gamma}$, then the leading order decay rates of
$\rho(x)$ and $m(x)$ are $\alpha$ and $2 \alpha$, respectively.  The
solution is a depression wave because $R(x) < 0$.  For subsonic
far-field conditions when $\alpha > \sqrt{\gamma}$, the leading order
decay rate is $\sqrt{\gamma}$, determined by the homogeneous solution.
Consequently, further analysis is required to deduce the polarity of
the solution.

\subsection{Hydraulic regime}
\label{sec:hydraulic-solution}
  
In this section, we investigate the asymptotic regime in which the
Wadati potential function is slowly varying \eqref{eq:32} by using the
hydraulic approximation \eqref{eq:22}.  A detailed analysis of the
hydraulic solutions is carried out in Appendix
\ref{sec:analys-hydr-appr} by analyzing the discriminant $D$ of the
quartic polynomial \eqref{eq:22} in $\rho$
\begin{equation}
  \label{eq:39}
  \begin{split}
    D(X) &= 4096 \rho_0^2 u_0^2 w(X)^4 \left ( 4\rho_0 - u_0^2 \right )
        \left ( \rho_0 + 2u_0^2 \right )^2K(X), \\
    K(X) &= 3(\rho_0 - u_0^2)^4 -4(2\rho_0 + u_0^2)(\rho_0^2 + 10 
        \rho_0 u_0^2 - 2 u_0^4)w(X)^2 + 6(\rho_0 - u_0^2)^2 w(X)^4 - w(X)^8 . 
  \end{split}
\end{equation}
When $u_0 = \sqrt{f_1(\rho_0)} \equiv 2 \sqrt{\rho_0}$,
$D(X) \equiv 0$.  The curves $u_0 = \sqrt{f_j(\rho_0)}$, $j = 2,3$
correspond to boundary values that are square roots of the two sole
real, positive zeros of the discriminant factor $K(X = 0) = 0$ in
eq.~\eqref{eq:39}.  These three zeros $u_0^2 = f_j(\rho_0)$ of the
discriminant occur at the peak of the Wadati potential function $w(X)$
when $X = 0$ and determine bifurcations in the existence and nature of
hydraulic solutions. Additional analysis of the discriminant
\eqref{eq:39} and the solutions of eqs.~\eqref{eq:9b} and
\eqref{eq:22} in Appendix \ref{sec:analys-hydr-appr} prove what
follows.

\begin{figure}
  \centering
  \includegraphics[scale = 0.8]{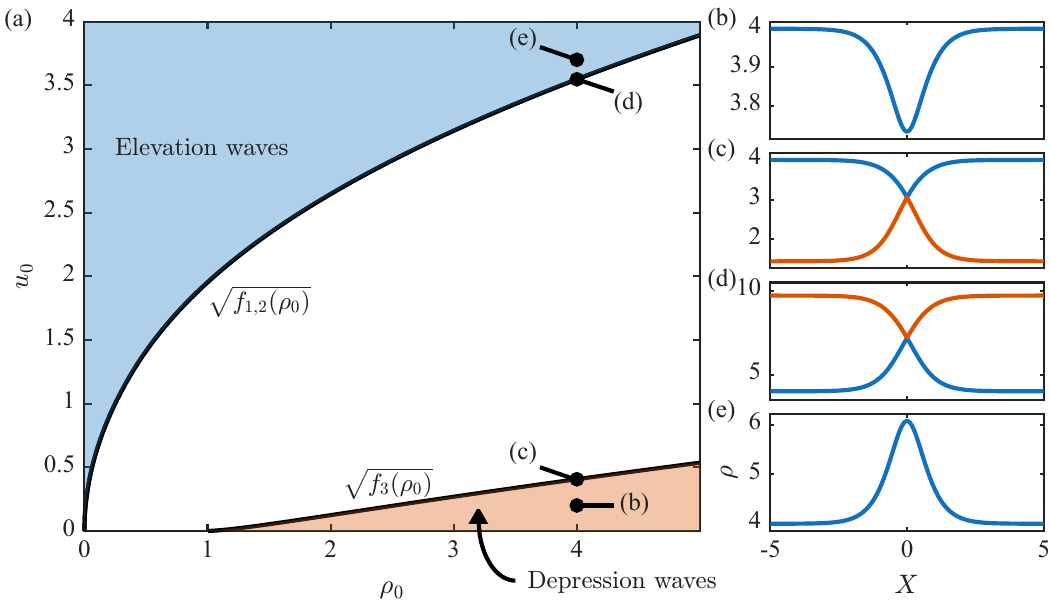}
  \caption{
  % \MH{Since the phase diagram is, I would say, the main
  %     result of this work, I think that it should be bigger.  Please
  %     shade and label the depression wave and elevation wave existence
  %     regions.  The density profiles could be half their width because
  %     they are just pulses.  Also recommend defining the upper curve
  %     continuously as $\sqrt{f_{1,2}(\rho_0)}$ rather than distinguish
  %     between 1 and 2.}\PS{figure is updated!} 
      (a) Hydraulic phase diagram depicting the
    existence of two-parameter families of depression and elevation
    homoclinic solutions of eq.~\eqref{eq:22} with boundary conditions
    $(\rho_0,u_0)$.  (b,c) Depression waves (blue).  (d,e) Elevation
    waves (blue).  (c,d) Kink and antikink heteroclinic solutions
    obtained by continuously joining elevation and depression waves at
    $X = 0$ when $u_0 = \sqrt{f_j(\rho_0)}$, $j = 1,2,3$.  The density
    profiles are for the Wadati potential function
    $w = \mathrm{sech}^2(X)$.}
  \label{fig:discriminant}
\end{figure}

The results are summarized in the $\rho_0$-$u_0$ phase diagram and
density profiles of Fig.~\ref{fig:discriminant}.  The zeros of $D$ are
the bifurcation curves.
% $u_0 = \sqrt{f_j(\rho_0)}$, $j = 1,2,3$ relating the boundary
% conditions in density and velocity in the phase diagram
% Fig.~\ref{fig:discriminant}(a).
The continuous curve
\begin{equation}
  \label{eq:40}
  u_0 = \sqrt{f_{1,2}(\rho_0)} \equiv
  \begin{cases}
    \sqrt{f_1(\rho_0)} , & 0 < \rho_0
                                                 \le \frac{1}{3}, \\
    \sqrt{f_2(\rho_0)}, & \rho_0 > \frac{1}{3},
  \end{cases}
\end{equation}
defines the lower bound $u_0 \ge \sqrt{f_{1,2}(\rho_0)}$, $\rho_0 > 0$
for the existence of elevation waves.
% solutions of the hydraulic equations \eqref{eq:22}, \eqref{eq:9b}
% that satisfy the correct boundary conditions \eqref{eq:47}.
An example is shown in Fig.~\ref{fig:discriminant}(e) for
$\rho_0 = 4$.  For $u_0 \le \sqrt{f_3(\rho_0)}$ and $\rho_0 > 1$,
there exist depression waves with % solution of the hydraulic
% equations \eqref{eq:22}, \eqref{eq:9b} that satisfies the correct
% boundary conditions \eqref{eq:47}.
an example in Fig.~\ref{fig:discriminant}(b).  Thus, we have
identified two-parameter families of elevation and depression waves.
For
\begin{equation}
  \label{eq:44}
  0 < u_0 < \sqrt{f_{1,2}(\rho_0)}, \quad 0 < \rho_0
  \le 1 \quad \mathrm{and} \quad
  \sqrt{f_3(\rho_0)} < u_0 < \sqrt{f_{1,2}(\rho_0)}, \quad
  \rho_0 > 1,
\end{equation}
there exist no homoclinic solutions.  % of the hydraulic equations
% satisfying the correct boundary conditions.  A detailed description
% of the existence regimes of elevation and depression waves is given
% in the App.~\ref{sec:analys-hydr-appr}.

When the velocity boundary condition coincides with the bifurcation
curve $u_0 = \sqrt{f_{1,2}(\rho_0)}$, the hydraulic equations admit an
elevation wave with boundary conditions $(\rho_0,u_0)$ and a
depression wave with different boundary conditions that meet at
$X = 0$ so that one-parameter families of smooth kinks and antikinks
can be constructed by piecewise combining the two waves at $X = 0$.
The kink satisfies the boundary condition $(\rho,u) \to (\rho_0,u_0)$
as $x \to -\infty$, the antikink as $x \to \infty$.  These scenarios
are depicted in Fig.~\ref{fig:discriminant}(c,d).

%%%%%%%%%%%%%%%%%%%%%%%%%%%%%%%%%%%%%%%%%%%%%%%

\subsection{Depression waves}
\label{sec:depression-waves}

The existence regions in the $\rho_0$-$u_0$ plane are determined by
performing continuation in $u_0$ in steps $\Delta u_0 = 0.0005$ from
the CI wave solution \eqref{eq:12} for fixed $\rho_0 \in (0,1]$ with
spacing $\Delta \rho_0 = 0.025$.  Accurate low velocity initial
guesses for $\rho_0>1$ with spacing $\Delta \rho_0 = 0.1$ are the
approximate solutions Eqs.~\eqref{eq:46}. Non-convergence is declared
when the number of iterations exceeds 20 for the Levenberg-Marquardt
and Newton bi-conjugate gradient solvers, and the residuals have not
decreased below $10^{-8}$.  Both computational methods are employed to
define the boundary of the existence region for depression homoclinic
orbits.  Evaluating eq.~\eqref{eq:56} at $x = 0$ with $u_m = u(0)$ and
$\rho_m = \rho(0)$, we obtain
\begin{equation}
\label{Verification}
-\frac{1}{2}u_0^2-\rho_0=-\frac{1}{2}u_m^2-\rho_m+\frac{1}{2}+\frac{1}{4\rho_m}\left(4\rho_m u_m^2-4\rho_m u_m+2\rho_m^2-4\rho_0 u_0^2-2\rho_0^2\right).
\end{equation}
During numerical continuation, the first $u_0$ when this condition is
not satisfied to within $10^{-8}$ is labeled $d(\rho_0)$, identifying
the existence region as $u_0 < d(\rho_0)$.  The depression wave
existence regions
\begin{equation}
  \label{eq:depression_existence}
  0 < u_0 < d(\rho_0), \quad \rho_0 > \rho_*,
\end{equation}
for $w(x) = {\rm sech}(x)$ and $w(x) = {\rm sech}^2(x)$ are shown in
the homoclinic summary sec.~\ref{sec:summ-homocl-solut} and are
qualitatively similar to the hydraulic theory prediction
$u_0 = \sqrt{f_3(\rho_0)}$.  The critical density $\rho_*$ in the
intercept $d(\rho_*) = 0$.

\begin{figure}
  \centering
  \includegraphics[scale = 0.4]{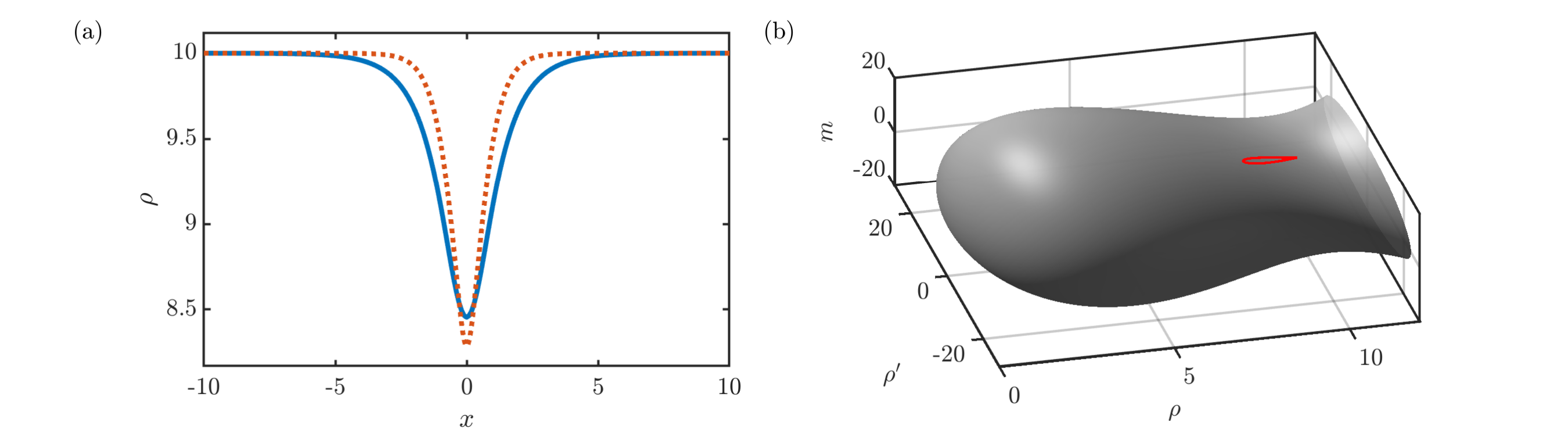}
  \caption{(a) Example depression waves (solid) compared with
    hydraulic theory (dashed).  (b) The zero energy surface
    (eq.~\eqref{eq:62}) defined by subsonic boundary conditions
    $\rho_0=10$, $u_0=1.1$, on which the homoclinic orbit for
    ${\rm sech}(x)$) (red) is shown.}
    \label{fig:depression_homoclinics_energy_surface}
\end{figure}
Figure \ref{fig:depression_homoclinics_energy_surface} displays
depression waves for $w(x)={\rm sech}(x)$ and their trajectory on the
zero energy surface \eqref{eq:62}.  We find that the homoclinic orbits
for $\rho(x)$ lie between the saddle
$(\rho,\rho',m) = (\rho_0,0,\sqrt{F(\rho_0))}$ and local maximum
$(\rho,\rho',m) = ((2u_0^2+\rho_0)/3,0,\sqrt{F((2u_0^2+\rho_0)/3)})$
of the energy surface.

Explicit, approximate depression waves can be obtained in various
asymptotic limits.  First, we consider their bifurcation from CI waves
\eqref{eq:12} for $u_0 = \epsilon$, $0 < \epsilon \ll 1$.  Inserting
the series expansion
\begin{equation}
  \label{eq:28}
  \rho = \rho_0 + \epsilon \rho_1(x) + \cdots, \quad u = w(x) +
  \epsilon u_1(x) + \cdots , \quad 0 < \epsilon \ll 1,
\end{equation}
into eq.~\eqref{eq:29}, we obtain at order $\mathcal{O}(\epsilon)$ the
homogeneous boundary value problem
% \begin{equation}
%   \label{Linearization-operator}
%   \mathbb{M}
%   \begin{bmatrix}
%     \rho_1 \\ u_1
%   \end{bmatrix}
%   =
%   \begin{bmatrix}
%     \frac{d^2}{dx^2}-4\rho_0&-4\rho_0 w(x)\\
%     w(x)\frac{d}{dx}& \rho_0\frac{d}{dx}
%   \end{bmatrix}
%   \begin{bmatrix}
%     \rho_1 \\ u_1
%   \end{bmatrix} =
%   \begin{bmatrix}
%     0 \\ 0
%   \end{bmatrix}, \quad
%   \begin{bmatrix}
%     \rho_1(x) \\ u_1(x)
%   \end{bmatrix} \to
%   \begin{bmatrix}
%     0 \\ 1
%   \end{bmatrix}, \quad |x| \to \infty .
% \end{equation}
% Equations \eqref{Linearization-operator} reduce to the third order
% scalar equation
\begin{equation}
  \label{Scalar-ODE-Bifurcation-from-CI-limit}
  u_1^{\prime\prime\prime}(x) + \left (-\frac{2w^{\prime}(x)}{w(x)}
  \right ) u_1^{\prime\prime}(x) + \bigg( \frac{2(w^{\prime})^2}{w^2}
  - \frac{w^{\prime\prime}}{w} - 4\rho_0+4w^2\bigg) u_1^{\prime}(x) +
  4w^{\prime}(x) w(x) u_1(x)=0,
\end{equation}
subject to $u_1(x) \to 1$ and $u_1'(x) \to 0$ as $|x| \to \infty$.
The density correction is
$\rho_1(x)=-\rho_0\int_{-\infty}^{x}u_1'(y)/w(y) dy$.

\begin{figure}
  \centering \includegraphics[scale = 0.75]{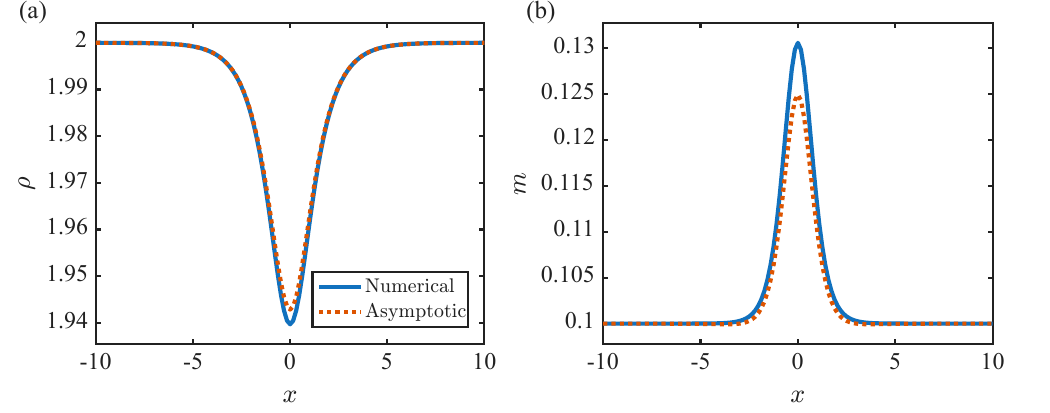}
  \caption{Small velocity depression wave bifurcating from constant
    intensity waves for $(\rho_0,u_0) = (2,0.05)$.}
  \label{fig:homoclinics_near_CI}
\end{figure}

For $w(x)={\rm sech}(x)$, the transformation $z=\tanh(x)$ of
\eqref{Scalar-ODE-Bifurcation-from-CI-limit} results in the quadratic
solution $u_1(\tanh^{-1}(z))=1+\frac{2}{4\rho_0-1}(1-z^2)$ so that
\begin{equation}
  \label{velocity-perturbation-1}
  u_1(x)=1+\frac{2}{4\rho_0-1}\mathrm{sech}^2(x) \quad \mathrm{and} \quad
  \rho_1(x)=-\frac{4\rho_0}{4\rho_0-1}{\rm sech}(x) , \quad \rho_0 \ne
  \frac{1}{4} .
\end{equation}
The two remaining linearly independent solutions of
eq.~\eqref{Scalar-ODE-Bifurcation-from-CI-limit} can be expressed in
terms of the associated Legendre functions $P_1^{2\sqrt{\rho_0}}(z)$,
$Q_1^{2\sqrt{\rho_0}}(z)$ that are unbounded in $x$ leaving 
\begin{equation}
  \label{eq:54}
  \rho(x) \sim \rho_0 - \frac{4\rho_0 u_0}{4\rho_0-1}{\rm sech}(x), \quad
  u(x) \sim w(x) + u_0 + \frac{2u_0}{4\rho_0-1}\mathrm{sech}^2(x),
  \quad 0 \le u_0 \ll 1,
\end{equation}
which approximates the two-parameter family of small velocity
depression waves bifurcating from CI waves, provided $\rho_0>1/4$. We
find that the numerical existence boundary
\eqref{eq:depression_existence} satisfies $\rho_* = 1/4$, i.e., the
special value $\rho_0 = 1/4$ is the $u_0$ intercept of the existence
curve $u_0 = d(\rho_0)$ for $w(x) = {\rm sech}(x)$. In
Fig.~\ref{fig:homoclinics_near_CI}, we favorably compare the
approximate and numerical solutions.  We have also numerically
computed the $u_0$ intercept for $w(x) = {\rm sech}^{n}(x)$ and find
that it is a decreasing function of $n$ from the value $\rho_0 = 1/4$
at $n = 1$.

Another asymptotic regime to analytically approximate depression waves
is the strongly nonlinear regime in which
$\rho = \mathcal{O}(\epsilon^{-1})$ and $w = W(x \epsilon^{-1/2})$
with $0 < \epsilon \ll 1$. We obtain an approximate solution by
inserting the ansatz
\begin{align}
  \label{eq:14}
  \rho(x) = \frac{1}{\epsilon} R_{0}(y) + R_1(y) + \cdots, \quad
  m(x) = \frac{1}{\epsilon} M_0(y) + M_1(y) + \cdots , \quad y = \frac{x}{\epsilon^{1/2}},
\end{align}
into eq.~\eqref{eq:29} with $\rho_0 = r_0/\epsilon$,
$r_0, ~u_0 = \mathcal{O}(1)$.  Then, the equations at
$\mathcal{O}(\epsilon^{-2})$ are
\begin{subequations}
  \label{eq:31}
  \begin{align}
    \label{eq:17}
    R_0'' - 6 R_0^2 + 8 r_0 R_0 - 2
    r_0^2 &= 0 , \\
    \label{eq:37}
    M_0' + W(y) R_0' &= 0 ,
  \end{align}
\end{subequations}
subject to the boundary conditions
\begin{equation}
  \label{eq:45}
  R_0(y) \to r_0, \quad M_0(y) \to r_0 u_0, \quad |y| \to \infty.
\end{equation}
The BVP \eqref{eq:31}, \eqref{eq:45} admits the unique, black
soliton depression wave solution
\begin{equation}
  \label{eq:30}
  R_0(y) = r_0\, \mathrm{tanh}^2 \left ( \sqrt{r_0} y \right )
  , \quad M_0(y) = \int_{y}^\infty W(\tilde{y})
  R_0'(\tilde{y})\,\mathrm{d}\tilde{y} + r_0 u_0 ,
\end{equation}
in which the density is zero at the origin $R_0(0) = 0$.

At the next order $\mathcal{O}(\epsilon^{-1})$, we have
\begin{subequations}
  \label{eq:55}
  \begin{align}
    \label{eq:67}
    R_1'' + V(y) R_1
    &= F_1(y), \\ 
    \label{eq:68}
    M_1' + W(y)R_1'
    &= 0 ,
  \end{align}
\end{subequations}
where
\begin{equation}
  \label{eq:13}
  \begin{split}
    V(y) = 4 r_0(-1 + 3 \mathrm{sech}^2(\sqrt{r_0}y)), \quad F_1(y) =
    4 W(y) M_0(y) + 4 (r_0-R_0(y)) u_0^2 . 
  \end{split}
\end{equation}
The boundary conditions for eq.~\eqref{eq:55} are
\begin{equation}
  \label{eq:51}
  R_j(y) \to 0, \quad M_j(y) \to 0, \quad |y| \to 0 .
\end{equation}

The self adjoint linear operator
$\frac{\mathrm{d^2}}{\mathrm{d}y^2} + V(y)$ in eq.~\eqref{eq:67}
subject to decaying boundary conditions admits a one-dimensional
kernel spanned by
\begin{equation}
  \label{eq:72}
  H(y) = \mathrm{tanh}(\sqrt{r_0}y) \mathrm{sech}^2(\sqrt{r_0}y) .
\end{equation}
One can check that $F_1(y)$ is even.  Consequently, its
$L^2(\mathbb{R})$ inner product with the odd kernel \eqref{eq:72} is
zero and equation \eqref{eq:67} for $R_1(y)$ is solvable.  Reduction
of order yields corrections to the density profile
\begin{equation}
  \label{eq:71}
  R_1(y) = H(y) \int^y \frac{1}{H^2(\tilde{y})} \int_{-\infty}^{\tilde{y}} H(z)
  F_1(z)\,\mathrm{d}z \,\mathrm{d}\tilde{y}, 
\end{equation}
where the integration constant of the outermost integral is selected
so that $R_1'(0) = 0$ and $R_1(y)$ is an even function.  The momentum
correction is
\begin{equation}
  \label{eq:73}
  M_1(y) = \int_y^\infty W(\tilde{y})
  R_1'(\tilde{y})\,\mathrm{d}\tilde{y} ,
\end{equation}
also an even function.  This process can be continued to arbitrary
order $\mathcal{O}(\epsilon^n)$.

\begin{figure}
    \centering
    \includegraphics[scale = 0.25]{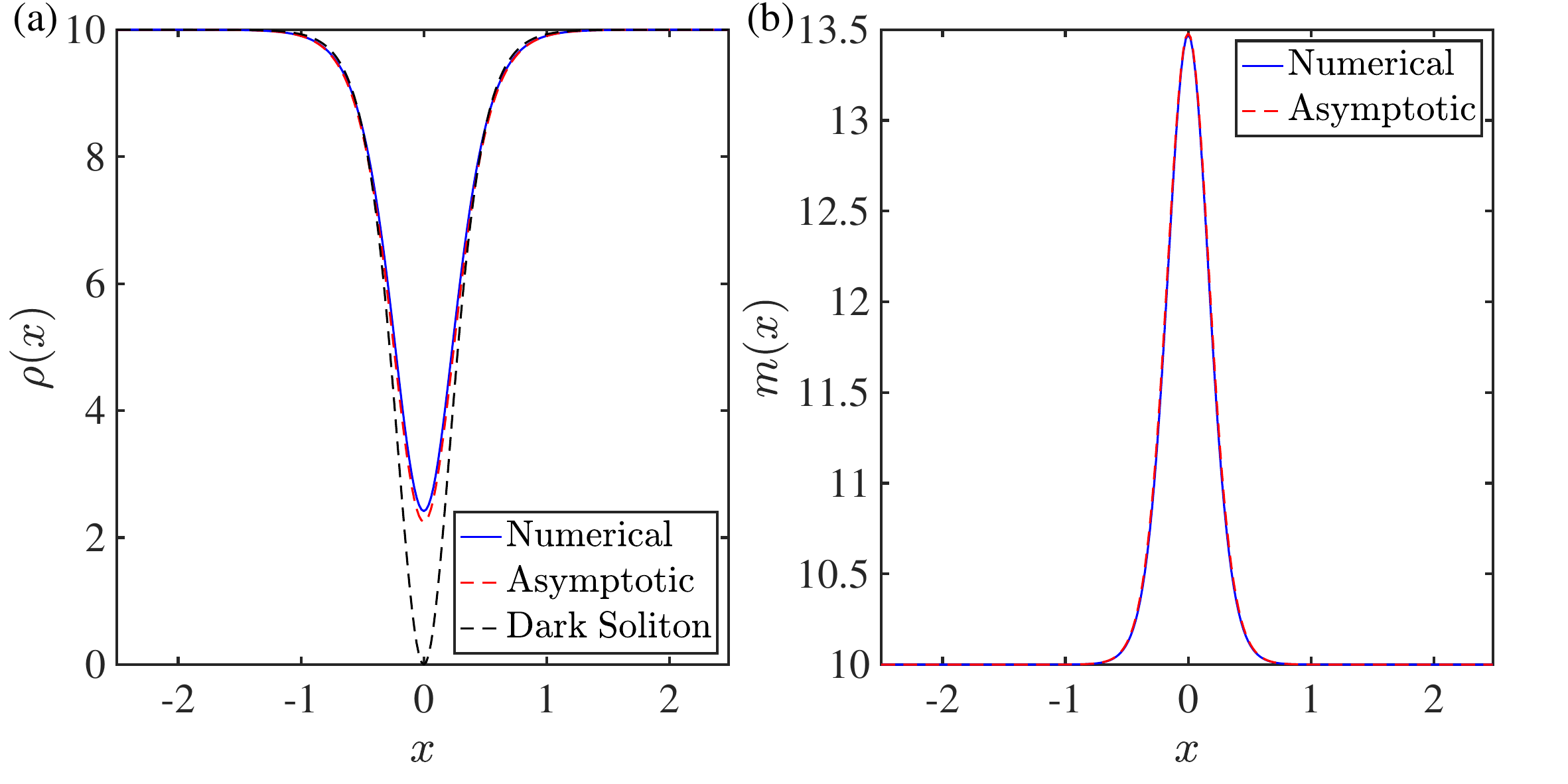}
    \caption{Depression wave in the large density $\rho_0 = 10$,
      $u_0 = 1$ regime: approximate (dashed red) and numerical (solid
      black).}
    \label{fig:DS-profile}
\end{figure}

% At higher orders, we obtain the system 
% \begin{equation}
%   \label{eq:75}
%   R_j'' + V(y) R_j = F_j(y), \quad M_j' + W(y) R_j' = 0, \quad j = 2,
%   3, \ldots,
% \end{equation}
% with boundary conditions $R_j(y) \to 0$, $M_j(y) \to 0$,
% $|y| \to \infty$, $j = 2,3,\ldots$ where the inhomogeneity $F_j(y)$
% depends on the solutions at lower orders in $\epsilon$.  Assuming
% $F_j(y)$ are even functions, these are similarly solved by
% \begin{equation}
%   \label{eq:76}
%   R_j(y) = H(y)\int^y \frac{1}{H^2(\tilde{y})}
%   \int_{-\infty}^{\tilde{y}} H(z)
%   F_j(z)\,\mathrm{d}z\,\mathrm{d}\tilde{y}, \quad M_j(y) =
%   \int_y^\infty W(\tilde{y})R_j'(\tilde{y})\,\mathrm{d}\tilde{y} ,
% \end{equation}
% subject to $R_j'(0) = 0$.  This calculation determines the asymptotic
% series in eq.~\eqref{eq:14} for the density $\rho(x)$ and shifted
% momentum $m(x)$ to all orders in $\epsilon$.

% One could verify that asymptotic ordering of the series \eqref{eq:14}
% is maintained by bounding $R_j(y)$ and $M_j(y)$.
% Instead, we
% explicitly compute the first two terms by
Choosing {$w(x) = \mathrm{sech}^2(\sqrt{\rho_0} x)$}, we obtain the
approximate solution
\begin{subequations}
  \label{eq:78}
  \begin{align}
    \label{eq:79}
    \rho(x) &\sim \rho_0 \,\mathrm{tanh}^2\left (\sqrt{\rho_0} x \right ) +
              \frac{1}{4}\, \mathrm{sech}^2 \left (\sqrt{\rho_0} x
              \right ) \left ( (1 + 2 u_0)^2 - 4
              u_0(1+u_0)\sqrt{\rho_0}\, x\,\mathrm{tanh}\left (
              \sqrt{\rho_0} x \right ) + \mathrm{tanh}^2\left (
              \sqrt{\rho_0} x \right ) \right ), \\
    m(x) &\sim \rho_0 u_0 + \frac{\rho_0}{2} \mathrm{sech}^4\left (
           \sqrt{\rho_0} x \right ), % + M_1\left ( \sqrt{\rho_0} x
                                     % \right ),
           \quad \rho_0 \to \infty .
  \end{align}
  % where
  % \begin{equation}
  %   \label{eq:80}
  %   \begin{split}
  %     M_1(y ) &= \frac{1}{15} u_0(1+u_0)\,\mathrm{sech}^2\left (
  %          y \right ) - \frac{1}{20}\left ( 5 +
  %                             12u_0(1+u_0) \right ) \,\mathrm{sech}^4\left (
  %          y \right ) + \frac{1}{6}\, \mathrm{sech}^6\left (
  %                             y \right )\\
  %                           &\quad + \frac{1}{30} u_0 (1+u_0)
  %                             y\, \mathrm{sech}^4\left (
  %          y \right )\tanh(y)\left (23 + 6 \,\cosh\left (
  %                             2y \right ) \right ) \\
  %     &\quad + \frac{1}{30} u_0(1+u_0) \left ( y \cosh\left (
  %       4 y \right )\,\mathrm{sech}^4\left (
  %          y \right ) \tanh\left (
  %          y \right ) - 8 \log \left ( 2 \cosh\left (
  %          y \right ) \right ) \right ) .
  %   \end{split}
  % \end{equation}
\end{subequations}
% Since $M_0 = |u_0|/\sqrt{\rho_0}\ll 1$, this solution is a subsonic
% depression wave, which agrees with the prediction from the linearized,
% far-field analysis in sec.~\ref{sec:linearized-far-field} since the
% Wadati potential function's decay rate $\alpha = \sqrt{\rho_0}$
% satisfies $\alpha < \sqrt{\gamma} \sim 2\sqrt{\rho_0}$ as
% $\rho_0 \to \infty$.
% Although in a different regime, since
% $\rho_0 \gg 1$, $u_0 = \mathcal{O}(1)$, the solution lies in
% the predicted existence region $u_0 \le \sqrt{f_3(\rho_0)}$ for
% hydraulic depression waves shown in Fig.~\ref{fig:discriminant}(a).
The solution \eqref{eq:78} is a perturbed dark soliton, exhibiting a
weak dependence on the Wadati potential through higher order
corrections.  The density and shifted momentum at the origin are
\begin{equation}
  \label{eq:81}
  \rho(0) \sim \frac{1}{4}(1 + 2u_0)^2 > 0, \quad m(0) \sim \rho_0 \left
    (u_0 + \frac{1}{2} \right ),
  % - \frac{1}{12} - \frac{4}{15}u_0(1+u_0)(2+\log 2) ,
  \quad \rho_0 \to \infty ,
\end{equation}
demonstrating that the Wadati potential function deforms the leading
order black soliton solution with zero density at $x = 0$ into a
depression wave with nonzero density at the origin. Figure
\ref{fig:DS-profile} favorably compares the asymptotic \eqref{eq:78}
and numerically computed solutions.

\subsection{Elevation waves}
\label{sec:elevation-waves}

\begin{figure}[H]
    \centering
    \includegraphics[width=\linewidth]{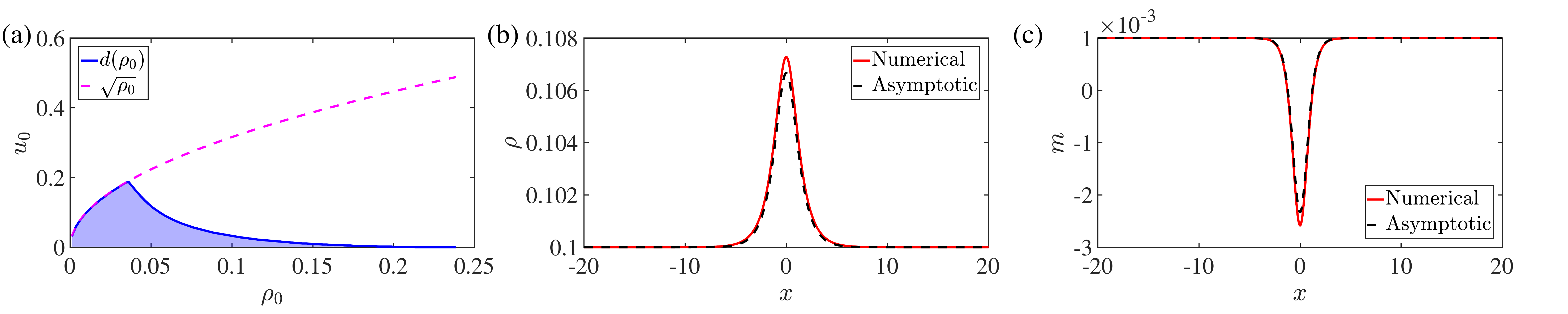}
    \caption{Elevation waves bifurcating from the constant intensity
      wave limit for $w(x) = {\rm sech}(x)$: (a) The numerical
      existence region is below the blue solid curve. The computed density (b)
      and shifted momentum (c) profiles for an elevation wave with
      $\rho_0=0.1$, $u_0=0.01$ together with comparisons to the asymptotic solution Eq.~\eqref{eq:54}.}%\MH{shade the
       % existence region in (a); compare with the asymptotic solution
        % in (b) and (c)}}
    \label{fig:elevation}
\end{figure}

The existence region and profile of elevation waves when
$w(x) = {\rm sech}(x)$ are shown in Fig.~\ref{fig:elevation}, where
the boundary of the existence region $d(\rho_0)$ is shown with the
sonic curve $u_0=\sqrt{\rho_0}$. For $\rho_0\lessapprox 0.036$, the
existence boundary is coincident with the sonic curve.
For $\rho_0\gtrapprox 0.36$, the admissible values of $u_0$ in the existence region sharply
diminish.
The asymptotic solution \eqref{eq:54} approximates elevation waves as
bifurcations from a CI wave when $w(x) = {\rm sech}(x)$,
$0 < \rho_0 < 1/4$, and $0<u_0 \ll 1$.  This approximate solution
compares favorably with the numerical solution depicted in
Fig.~\ref{fig:elevation}(b,c).

Another asymptotic regime that exhibits elevation waves is that of
small density with $\rho(x) = \mathcal{O}(\epsilon)$ and
$u_0 = \mathcal{O}(1)$ for $0 < \epsilon \ll 1$, $x \in \R$.
Inserting the asymptotic expansion
\begin{equation}
  \label{eq:2}
  \rho(x) = \epsilon R_1(x) + \epsilon^2 R_2(x) + \cdots, \quad m(x)
  = \epsilon M_1(x) + \epsilon^2 M_2(x) + \cdots,
\end{equation}
into eq.~\eqref{eq:29} with $\rho_0 = \epsilon$ and $u_0 > 0$ an order
one quantity as $\epsilon \to 0$ and equating like powers of
$\epsilon$, we obtain the inhomogeneous linear system of equations at $\mathcal{O}(\epsilon)$
\begin{align}
  \label{eq:5}
  R_1'' + 4 u_0^2 R_1  - 4 w
  M_1 = 4 u_0^2, \quad M_1' = -w R_1' ,
\end{align}
subject to the boundary conditions
\begin{equation}
  \label{eq:53}
  R_1(x) \to 1, \quad M_1(x) \to u_0, \quad |x| \to \infty.
\end{equation}
Equations \eqref{eq:5} admit the particular solution $R_1(x) = 1$,
$M_1(x) = 0$.  This solution approximates a constant intensity wave
when $u_0 = 0$.  But, when $u_0 > 0$, the boundary conditions
\eqref{eq:53} are not satisfied so we must add a nonzero homogeneous
solution of eq.~\eqref{eq:5} to the particular solution that enforces
the boundary conditions. 

\begin{figure}[H]
    \centering
    \includegraphics[scale = 0.8]{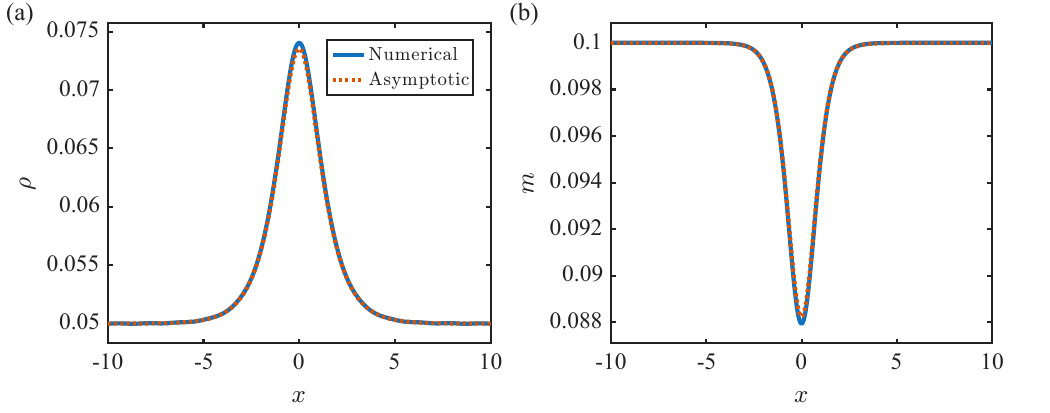}
    \caption{Comparison of the asymptotic solution \eqref{eq:2}
      and\eqref{eq:26} with the numerical solution for elevation waves
      with small density $\epsilon = \rho_0 = 0.05$, and velocity
      $u_0 = 2$ boundary conditions.}
    \label{fig:small_rho_homoclinics}
\end{figure}

When $w(x) = \mathrm{sech}(x)$, the system \eqref{eq:5} subject to
\eqref{eq:53} is solved exactly with
\begin{equation}
  \label{eq:26}
  R_1(x) = 1 + \frac{4 u_0}{1 + 4 u_0^2} \mathrm{sech}(x), \quad
  M_1(x) = u_0 \left ( 1 - \frac{2 \mathrm{sech}^2(x)}{1 + 4 u_0^2}
  \right ) .
\end{equation}
This approximate solution is only weakly affected by nonlinearity in
the higher order terms of \eqref{eq:2}.  Since the far-field flow
$u_0 \gg \rho_0 = \epsilon > 0$ is deep in the supersonic regime, the
solution lies on the disconnected ellipsoid of the zero energy surface
\eqref{eq:62}; see Fig.~\ref{fig:zero_energy_surface}.  A favorable
comparison of the asymptotic solution \eqref{eq:26} with the numerical
solution is shown in Fig.~\ref{fig:small_rho_homoclinics}.  Although
not visible in Fig.~\ref{fig:small_rho_homoclinics}, a careful
examination of the numerical solution reveals small amplitude
oscillations in the density.  These resonant waves will be explored in
the next section.

%%%%%%%%%%%%%%%%%%%%%%%%%%%%%%%%%%%%%%%%%%%%%%%%%%%%%%%%%%%%%%%%%%%%%%
\subsection{Generalized elevation waves}
\label{sec:elev-homocl-orbits}

In the supersonic regime $u_0>\sqrt{\rho_0}$, hydraulic theory
predicts elevation waves (cf.~Fig.~\ref{fig:discriminant}).  These
solutions are the leading order solutions to the singularly perturbed
ODE \eqref{eq:24}.  Linearizing eq.~\eqref{eq:24} for small
$\zeta = \rho - \rho_0$ results in
\begin{equation}
  \label{eq:77}
  \epsilon^2 \zeta'' - 4 \mu \zeta = 4 w m, \quad m' = - w \zeta' .
\end{equation}
As shown in sec.~\ref{sec:linearized-far-field}, the general solution
of eq.~\eqref{eq:77} obeys
\begin{equation}
  \label{eq:82}
  \zeta(X) \sim A e^{-\alpha X} + c_- e^{-2 i 
    \sqrt{u_0^2-\rho_0} X/\epsilon} + c_+ e^{2 i \sqrt{u_0^2 -
      \rho_0}X/\epsilon}, \quad X \to \infty ,
\end{equation}
for $w(X) = \mathcal{O}(e^{-\alpha X})$.  Since the far-field flow is
supersonic, stationary, small amplitude density waves with wavenumber
\begin{equation}
  \label{eq:83}
  k_\epsilon = \frac{2 \sqrt{u_0^2 - \rho_0}}{ \epsilon},
\end{equation}
are resonant.  In contrast, the shifted momentum $m(X)$ decays to
$\rho_0 u_0$ (cf.~eq.~\eqref{eq:11}).  Thus, when resonant waves are
present, elevation waves in density that decay to $\rho_0$ do not
exist. Stationary pulses on an oscillatory background are called
generalized solitary waves
\cite{pomeau_structural_1988,akylas_solitary_1992,akylas1995short,yang_nonlinear_2010,grimshaw2010exponential}.
Methods to analyze generalized solitary waves typically involve
exponential asymptotics, which we will utilize here.

An analytically tractable regime is large velocity $u_0$ in which we
define $0 < \varepsilon = 1/u_0 \ll 1$ with $\rho_0 = \mathcal{O}(1)$.
To this end, we introduce the expansion
\begin{equation}
  \label{Regular-perturbation-series}
  \rho(x) = \rho_0 + \varepsilon \rho_1(x) + \varepsilon^3
  \rho_3(x)+ \cdots,\;m(x) = \frac{\rho_0}{\varepsilon} + \varepsilon
  m_1(x)+ \varepsilon^3 m_3(x)+ \cdots,
\end{equation}
where $\rho_j(x)$ and $m_j(x)$ for $j \ge 1$ all satisfy zero boundary
conditions.  Inserting the expansion
\eqref{Regular-perturbation-series} into eq.~\eqref{eq:29} and
equating like powers of $\varepsilon$ results in the leading order
solution
\begin{equation}
  \label{eq:84}
  \rho_1(x) = \rho_0 w(x) , \quad m_1(x) = - \frac{\rho_0}{2} w(x)^2 .
\end{equation}
The corrections at each of the next algebraic orders involve only
derivatives and powers of $\rho_1(x)$ and $m_1(x)$.  Consequently, the
expansion \eqref{Regular-perturbation-series} remains localized in
$x$.  In order to estimate the non-decaying oscillatory terms from the
far-field analysis of eq.~\eqref{eq:16}, we let $\zeta(x) =
\rho(x)-\rho_0$ and $\eta(x) = m(x) - \rho_0/\varepsilon$ and
linearize eq.~\eqref{eq:29} to obtain
\begin{equation}
\label{ODE-linear-singular}
\frac{\varepsilon^2}{4} \zeta''(x)+ \zeta(x)
=\varepsilon \rho_0 w(x),
\end{equation}
where we have neglected the term proportional to
$\varepsilon \eta(x)w(x)$ as being of higher order because $\eta(x)$
is non-oscillatory in the far-field (cf.~eq.~\eqref{eq:11}).  Using
variation of parameters, we obtain the particular solution
\begin{equation}
  \label{solution-ODE-linear-singular}
  \zeta_p(x)= 2 \rho_0 \int_x^\infty \sin(k_\varepsilon(y-x)) w(y)
  \,\mathrm{d} y, \quad k_\varepsilon = \frac{2}{\varepsilon}
\end{equation}
where $k_{\varepsilon}=2|u_0|$ is approximately equal to the resonant wavenumber $k=2\sqrt{u_0^2-\rho_0}$ in the regime $u_0\gg \rho_0$.
The general solution of eq.~\eqref{ODE-linear-singular} is thus
\begin{equation}
  \label{eq:85}
  \zeta(x) = \zeta_p(x) + c \cos(k_\varepsilon x) + c_0
  \sin(k_\varepsilon x) .
\end{equation}
Requiring an even solution by setting $\zeta'(0) = 0$, we obtain
\begin{equation}
  \label{eq:86}
  c_0 = - \frac{\zeta_p'(0)}{k_\varepsilon} = \rho_0 \int_{-\infty}^\infty
  \cos(k_\varepsilon y) w(y)\,\mathrm{d} y ,
\end{equation}
leaving $c \in \mathbb{R}$ as a free parameter.  Note that
$c_0 = \rho_0 \mathrm{Re} \, \hat{w}(k_\varepsilon)$ where
$\hat{w}(k) = \int_{-\infty}^\infty w(x) e^{ikx}\,\mathrm{d}x$ is the
Fourier transform of $w(x)$. The far-field asymptotics are therefore
\begin{equation}
  \label{eq:88}
  \zeta(x) \sim c \cos(k_\varepsilon x) \pm c_0 \sin(k_\varepsilon x)
  = R_0 \cos(k_\varepsilon x - \mathrm{sgn}(x) \sigma )
  \quad x \to \pm \infty ,
\end{equation}
with amplitude and phase shift
\begin{equation}
  \label{eq:89}
  R_0 = \sqrt{c^2 + c_0^2}, \quad 2 \sigma = 2 \tan^{-1}(c_0/c) ,
  \quad c \in \R .
\end{equation}
We can eliminate the free parameter $c = c_0 \cot \sigma$ to obtain
the amplitude-phase shift relation
\begin{equation}
  \label{eq:90}
  R_0 = c_0 \csc \sigma , \quad \sigma \in (0,\pi/2] .
\end{equation}
For example, if $w(x) = {\rm sech}(x)$, then $c_0 = \pi \rho_0 {\rm
  sech}(\pi/\varepsilon)$.

Thus, to extend the regular perturbation expansion \eqref{Regular-perturbation-series} so as to incorporate the exponentially small periodic tails, we write
\begin{equation}
\label{eq:91}
\rho(x) \sim \rho_0 + \zeta(x), \qquad
m(x) \sim -\int_{-\infty}^{x} w(y)\zeta'(y)dy,
\end{equation}
where $\zeta(x)$ is a three-parameter family of approximate solutions
characterized by ${\sigma,\rho_0,\varepsilon}$ (see
Eq.~\eqref{eq:85}).
%\MH{Is this a three parameter family $(\rho_0,\varepsilon,\sigma)$ of
 % solutions?}

\begin{figure}
  \centering
  \includegraphics[width=\linewidth]
  {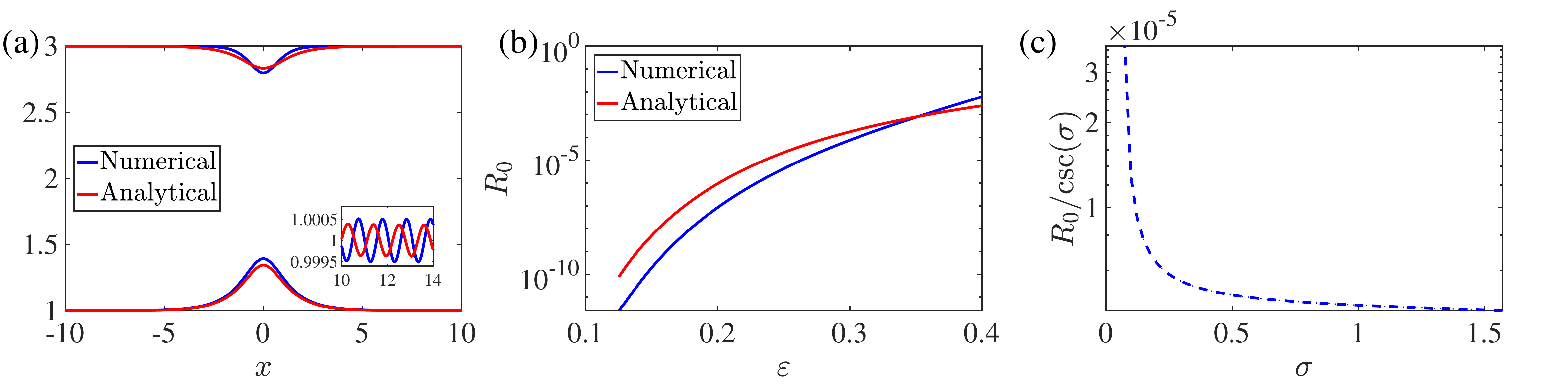}
  \caption{Generalized solitary waves for $\rho_0=1$, $w(x)=\sech(x)$,
    and variable $u_0 = \varepsilon^{-1}$. (a) Density (lower pair of
    curves) and shifted momentum (upper pair of curves) profiles for
    $\varepsilon =1/3$ and {$\sigma = \pi/2$}. The inset magnifies the
    far-field density oscillations.  (b) Numerical (blue) and
    analytical (red) tail amplitudes versus $\varepsilon$ for
    $\sigma=\pi/2$.  (c) Measured tail amplitude $R_0$ divided by
    $\csc \sigma$ from numerical computations as a function of
    $\sigma$ for $\varepsilon = 0.25$. The scaled amplitude remains
    approximately constant, as predicted by eq.~\eqref{eq:90}, for
    $\sigma\gtrsim 0.5$.}
    \label{fig:Resonant_homoclinics_non_eps}
\end{figure}
To corroborate these analytical findings, we numerically compute
generalized elevation waves.  Figure
\ref{fig:Resonant_homoclinics_non_eps} shows a comparison between the
numerical and analytically predicted generalized elevation wave
profile and the tail amplitude $R_0$, exhibiting good agreement in the
asymptotic regime of interest. In particular, the analytical
prediction for the tail amplitude $R_0$ remains within one order of
magnitude of the corresponding numerical value throughout the range of
$\varepsilon$ considered despite the linearization assumption in the
approximation eq.~\eqref{ODE-linear-singular}.

\subsection{Summary of homoclinic solutions}
\label{sec:summ-homocl-solut}

\begin{figure}
  \centering
  \includegraphics[width=0.35\linewidth]{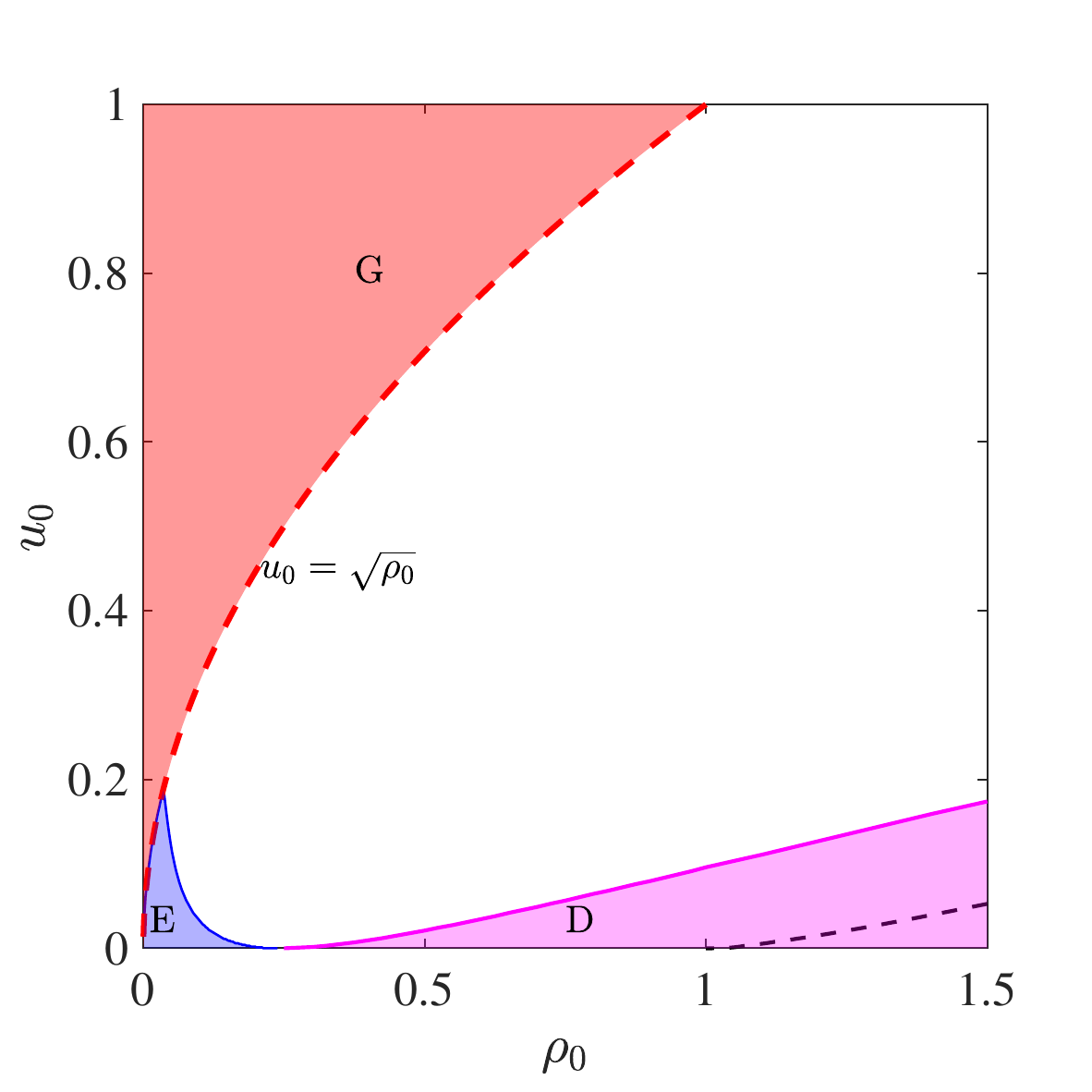}
  \caption{Existence diagram for elevation (E, blue), generalized
    elevation (G, orange), and depression (D, magenta) waves for
    $w(x)={\rm sech}(x)$.}
    \label{fig:homoclinic_sech}
\end{figure}

Figure \ref{fig:homoclinic_sech} is the existence diagram for
generalized elevation, elevation, and depression waves when
$w(x) = {\rm sech}(x)$.  By contrasting this figure with
Fig.~\ref{fig:discriminant}(a) from hydraulic theory, we see that
there are qualitative similarities: depression waves for low $u_0$ and
$\rho_0$ above a critical value, generalized elevation waves for
supersonic velocities.  On the
other hand, the existence of elevation waves in the low
density/velocity regime and the structure of generalized elevation
waves in the supersonic regime are completely missed by hydraulic
theory.  Qualitatively similar existence regions to
Fig.~\ref{fig:homoclinic_sech} can be obtained for other Wadati
potentials $w(x)$.  We have examined existence for
$w(x) = {\rm sech}^2(x)$ in detail and find it to be very similar to
the case $w(x) = {\rm sech}(x)$.

In the hydraulic regime, depression waves at the existence boundary
exhibit a subsonic far-field flow that is accelerated to supersonic
speeds at its center. For narrower potentials, depression waves near
the existence boundary exhibit supersonic flow in the vicinity of the
density minimum.  Elevation waves are subsonic throughout.  Supersonic
far-field flows result in generalized elevation waves that resonate
with stationary, oscillatory small amplitude waves.

\section{Heteroclinic solutions}
\label{sec:heter-solut}

In this section, we shift our attention to heteroclinic solutions of
eq.~\eqref{eq:29} subject to the boundary conditions \eqref{eq:3} when
$u_+ \ne u_-$.  Applying these boundary conditions to
eqs.~\eqref{eq:56} and \eqref{eq:57} results in the following
necessary conditions for the existence of heteroclinic solutions
\begin{subequations}
  \label{eq:87}
  \begin{align}
    \label{Eqn-mu}
    & -\mu = \frac{1}{2}u_{-}^2+\rho_-=\frac{1}{2}u_{+}^2+\rho_+\\
    \label{Eqn-A}
    & A =\rho_- u_-^2+\frac{1}{2}\rho_-^2=\rho_+ u_+^2+\frac{1}{2}\rho_+^2
  \end{align}
\end{subequations}
The system of equations \eqref{eq:87} admit two two-parameter families of solutions
\begin{align}
  \label{eq-hetero-1}
  &\textrm{Type-I:} \quad \rho_-=\rho_+,\;\;u_-=-u_+, \quad \rho_+ >
    0, \quad u_+ \in \mathbb{R}, \\
  \label{eq-hetero-2}    
  &\textrm{Type-II:} \quad
    \rho_-=\frac{1}{3}\left(2u_{+}^2+\rho_+\right),\;\; u_-=\pm 
    \frac{1}{\sqrt{3}}(4\rho_+-u_+^2)^{1/2} , \quad \rho_+ > 0, \quad
    |u_+| \le 2\sqrt{\rho_+},
\end{align}
which relate the left, $-$ state to the right, $+$ state.  We refer to
the two heteroclinic families as Type-I \eqref{eq-hetero-1} and
Type-II \eqref{eq-hetero-2}.  While Type-I heteroclinics exhibit the
same far-field Mach number $M_\pm = |u_\pm|/\sqrt{\rho_\pm}$, Type-II
heteroclinics generally differ.  % We find that $M_+ = M_-$ when
% \begin{equation}
%   \label{eq:33}
%   \rho_+ = \frac{u_+^2(3+4 u_+^2+3 \sqrt{1+24 u_+^2})}{2(12-u_+^2)} ,
% \end{equation}
% both of which are subsonic when $u_+ > 1$ and supersonic when
% $0 < u_+ < 1$.  Equation \eqref{eq:33} can be inverted to obtain one
% real, positive branch of $u_+ = u_+(\rho_+)$.

% \MH{Plot left versus right Mach number}

The second symmetry in eq.~\eqref{symmetries} implies that
heteroclinic solutions, if they exist, have both polarities: positive
(increasing density) and negative (decreasing density). However,
unlike their homoclinic counterparts, these heteroclinic solutions
$\psi(x,t)=\sqrt{\rho(x)}e^{i\int^{x} u(s) ds + i \mu t}$ are not
PT-symmetric. We refer to Type-II heteroclinics with negative polarity
($\rho_->\rho_+$) as antikinks and those with positive polarity
($\rho_+>\rho_-$) as kinks.

\subsection{Type-I heteroclinic solutions bifurcating from constant
  intensity waves}
\label{Heteroclinic-CI-solutions}

We begin our analysis of Type-I kinks for the Wadati potential
$w(x) = {\rm sech}(x)$ that bifurcate from the CI wave
$\rho(x) = \rho_0$, $u(x) = w(x)$.  The asymptotic expansion
\eqref{eq:28} in the small velocity $|u(x) - w(x)| \ll 1$, near-CI
wave limit, satisfies the linear homogeneous equation
\eqref{Scalar-ODE-Bifurcation-from-CI-limit} at first order.  We
showed that the density value $\rho_0 = 1/4$ is a bifurcation point
between approximate elevation and depression waves where
eq.~\eqref{Scalar-ODE-Bifurcation-from-CI-limit} admits one linearly
independent, bounded solution for fixed $\rho_0$.  At the special
value $\rho_0 = 1/4$, eq.~\eqref{Scalar-ODE-Bifurcation-from-CI-limit}
admits two linearly independent, bounded solutions that result in the
general solution
\begin{equation}
  \rho_1(x) = -\frac{c}{2} {\rm sech}(x) + c_0 x {\rm sech}(x), \quad
  u_1(x) = c {\rm sech}^2(x) + 2c_0({\rm tanh}(x) + x {\rm
    sech}^2(x)) , \quad c,c_0 \in \R .
\end{equation}
Since $u_1(x) \to \pm 2 c_0$ as $x \to \pm \infty$, we set $c_0 =
-1/2$ without loss of generality and obtain the approximate Type-I
kink
\begin{equation}
  \label{Formula-same-density}
  \rho(x) \sim 1/4 - \frac{\epsilon}{2} (c\, {\rm sech}(x)+x{\rm
    sech}(x) ), \quad u(x) \sim {\rm sech}(x)+ \epsilon (c\, {\rm
  sech}^2(x) - \tanh(x) - x{\rm sech}^2(x)), \quad 0 < \epsilon \ll 1,
\end{equation}
and $c \in \R$.  Because the density is not even, this solution
represents a symmetry-breaking bifurcation from the CI wave.  The
boundary conditions $u_\pm = \mp \epsilon$ and $\rho_0 = 1/4$ are
independent of $c$, so we have identified a two-parameter family of
Type-I kinks depending on $\epsilon > 0$ and $c \in \R$.

\begin{figure}
  \centering
  \includegraphics[width=\linewidth]{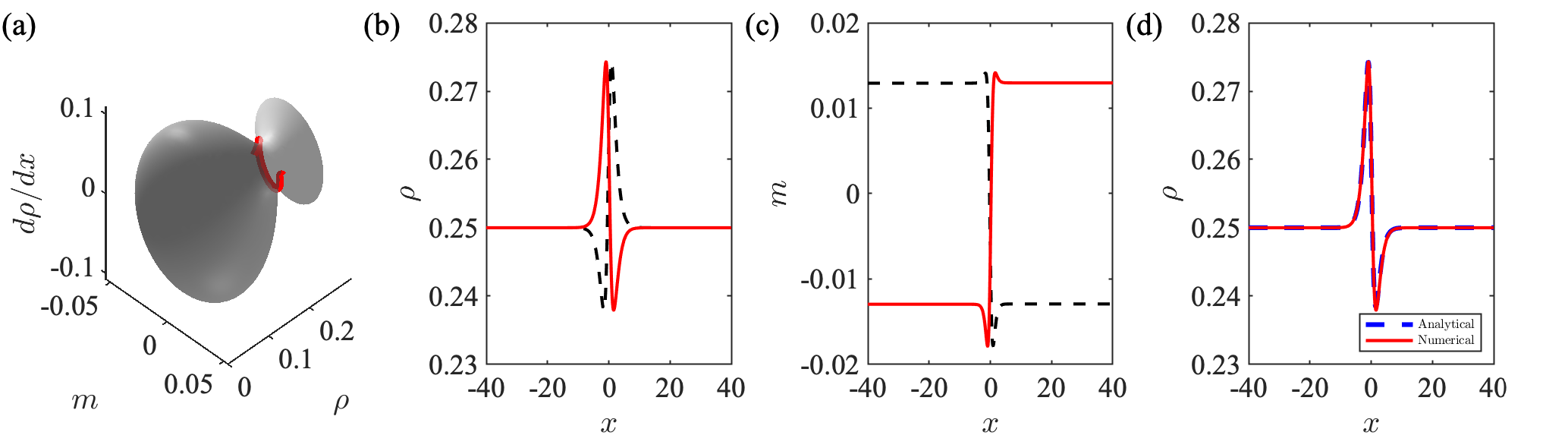}
  \caption{Type-I antikink for $w={\rm sech}(x)$. (a) Phase portrait
    of the antikink on the zero energy surface's neck. The antikink
    (solid red) and kink (dashed black) density (b) and shifted
    momentum (c).  (d) Comparison of numerical (solid red) and
    asymptotic (dashed black) solutions.}
  \label{fig:hetero1}
\end{figure}

We use the Wadati potential function $w(x)={\rm sech}(x)$ and an
initial guess \eqref{Formula-same-density} to compute a Type-I kink
for $\rho_0=0.25$ in the small velocity limit
$u_+=-u_{-}\approx 0.0519$ in Fig.~\ref{fig:hetero1}. The numerical
solution is depicted on the neck of the zero energy surface
\eqref{eq:62} in Fig.~\ref{fig:hetero1}(a). The solution is compared
with the approximate solution \eqref{Formula-same-density} by
performing a nonlinear least-squares fit to determine the parameters ({$\epsilon \approx 0.0504$\ and $c \approx -0.9623$}), displaying good agreement in
Fig.~\ref{fig:hetero1}(d).

\begin{figure}
  \centering
  \includegraphics[width=\linewidth]{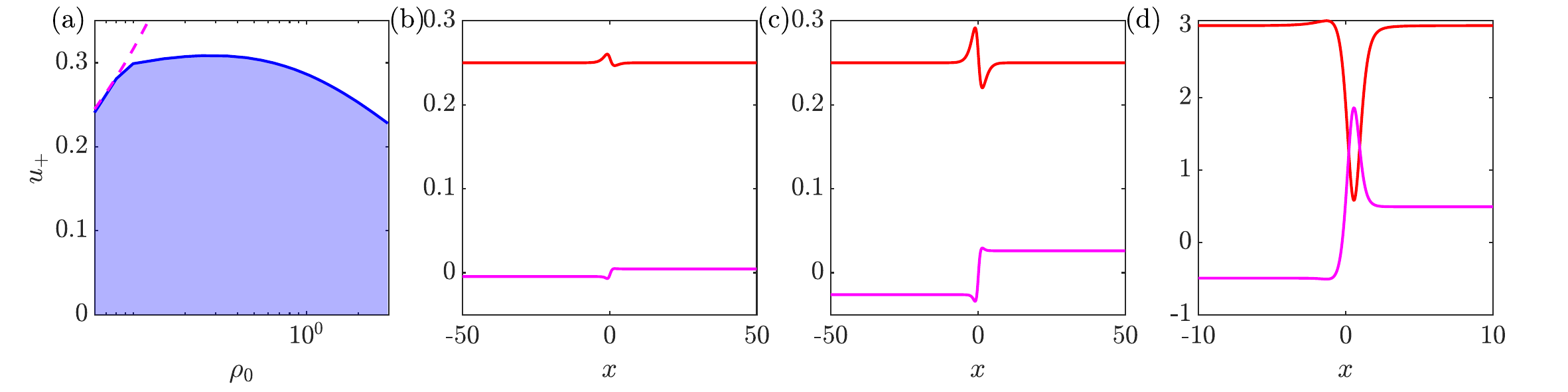}
  \caption{Two-parameter family of Type-I heteroclinics computed for
    $w(x)={\rm sech}(x)$. (a) Numerical existence boundary (solid
    blue) and the sonic curve $u_+=\sqrt{\rho_0}$ (dashed magenta).
    (b, c, d) Density (red) and shifted momentum (magenta) of Type-I
    antikinks {for $(\rho_0,u_+) = (0.25,0.0184),\, (0.25,0.1046), \, (3,1.974)$}.}
    \label{fig:hetero2}
\end{figure}

We also compute a two-parameter family of Type-I antikinks.  The
results are summarized in Fig.~\ref{fig:hetero2} where the boundary of
the numerical existence curve approaches the sonic curve
$u_0=\sqrt{\rho_0}$ at $u_0\approx 0.088$, signaling a likely
bifurcation there. This will be confirmed in the next section on
Type-II kinks. Performing velocity and density continuation from the
solution in Fig.~\ref{fig:hetero2} (b), we obtain solutions in
Fig.~\ref{fig:hetero2}(c,d), illustrating the two-dimensionality of
the solution family.

\subsection{Type-II heteroclinic solutions: hydraulic regime}
\label{sec:slowly-vary-potent-Type-II}

The discussion in Sec.~\ref{sec:hydraulic-solution} identified two
scenarios leading to the formation of a Type-II heteroclinic solution
using hydraulic theory. The kink is generated by the merger of two
roots representing homoclinic solutions along the curve
$u_0=\sqrt{f_3(\rho_0)}$, $\rho_0 > 1$ or the curve
$u_0=\sqrt{f_2(\rho_0)}$, $\rho_0\geq 1/3$. This root annihilation
is reminiscent of a saddle-node bifurcation, also pointed out in the
conservative case with an attractive potential
\cite{leszczyszyn_transcritical_2009}.

We now discuss, in further detail, Type-II kinks in the hydraulic
regime that satisfy the algebraic eqs.~\eqref{eq:22} and
\eqref{eq:9b}.  To obtain the kink curve of existence directly, we
define the density $\rho_m = \rho(0)$ and velocity $u_m = u(0)$ at the
Wadati potential peak and solve eqs.~\eqref{eq:22} and
\eqref{eq:9b} for
\begin{equation}
  \label{eq:10}
  \mu= \frac{1}{2}-\frac{1}{2}u_m^2-\rho_m,\quad A = \rho_m
  u_m^2-\rho_m u_m+\frac{1}{2}\rho_m^2.
\end{equation}
Differentiating eq.~\eqref{eq:22}, evaluating it at $X = 0$ and using
$w'(0) = 0$, we obtain
\begin{equation}
  \label{Sextic}
  8 \rho_m(u_m-1)(\rho_m - u_m^2) \rho'(0) = 0 .
\end{equation}
For kinks, we assume that $\rho'(0) \ne 0$ so that eq.~\eqref{Sextic}
implies either $u_m = 1$ or $u_m = \pm \sqrt{\rho_m}$ at the peak of
$w(x)$.  For the Type-II kinks satisfying $u_+ = \sqrt{f_2(\rho_+)}$,
$\rho_+ \ge 1/3$ or Type-II antikinks satisfying
$u_- = \sqrt{f_3(\rho_-)}$, $\rho_- > 1$, we find that they exhibit
the sonic condition $u_m = \sqrt{\rho_m}$ at $X = 0$.  This means that
a Type-II kink is subsonic for $X < 0$, supersonic for $X > 0$ and
vice-versa for an antikink.

\subsection{Type-II heteroclinics: numerical solutions}
\label{sec:numerical-solutions-1-Type-II}

\begin{figure}
    \centering
    \includegraphics[width=\linewidth]{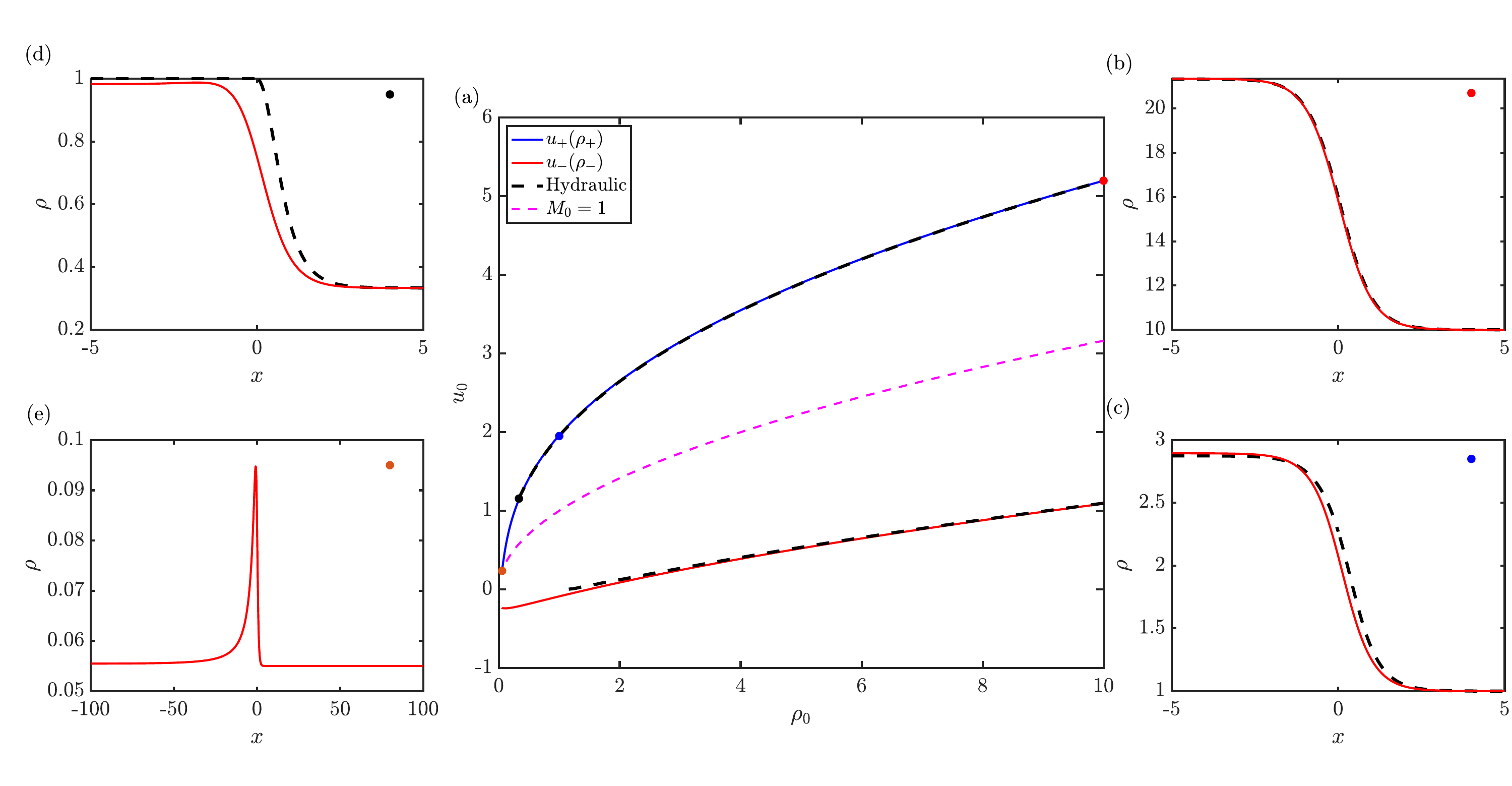}
    \caption{Type-II antikinks for $w(x)={\rm sech}^2(x)$.  (a)
      Existence curves $u_+(\rho_+)$ (solid blue) and
      $u_{-}(\rho_{-})$ (solid red).  Hydraulic theory curves
      $u_+ = \sqrt{f_2(\rho_+)}$ and $u_- = \sqrt{f_3(\rho_-)}$
      (dashed black). Red density profiles are denoted by a colored
      dot in (a) for (b) $\rho_+=10$ (red), (c) $\rho_+=1$ (blue), (d)
      $\rho_+=1/3$ (black), (e) $\rho_+\approx 0.0558$ (red) and
      corresponding hydraulic theory profiles for (b)-(d) (dashed
      black).}
    \label{fig:heteroclinic-solutions}
\end{figure}
Beyond the calculations in the hydraulic limit, we numerically compute
heteroclinic solutions of eq.~\eqref{eq:29}.  Figure
\ref{fig:heteroclinic-solutions} display the existence curves
$u_\pm(\rho_\pm)$ and example computed densities for Type-II
antikinks.  The numerical curves are close to their hydraulic theory
counterparts $\sqrt{f_2(\rho_+)}$ and $\sqrt{f_3(\rho_-)}$.  The
numerical existence curves continue further into the low density
regime before terminating at $\rho_{\pm}\approx 0.0558$.

%\MH{Recommend adding a plot of $M_+ = u_+(\rho_+)/\sqrt{\rho_+}$ and
 % $M_- = u_-(\rho_+)/\sqrt{\rho_-(\rho_+)}$ for antikinks to assess
  %whether one is always super/subsonic.  This is the case of hydraulic
  %solutions so it is natural to ask if it is so in general.}

\begin{figure}
    \centering
    \includegraphics[width=0.42\linewidth]{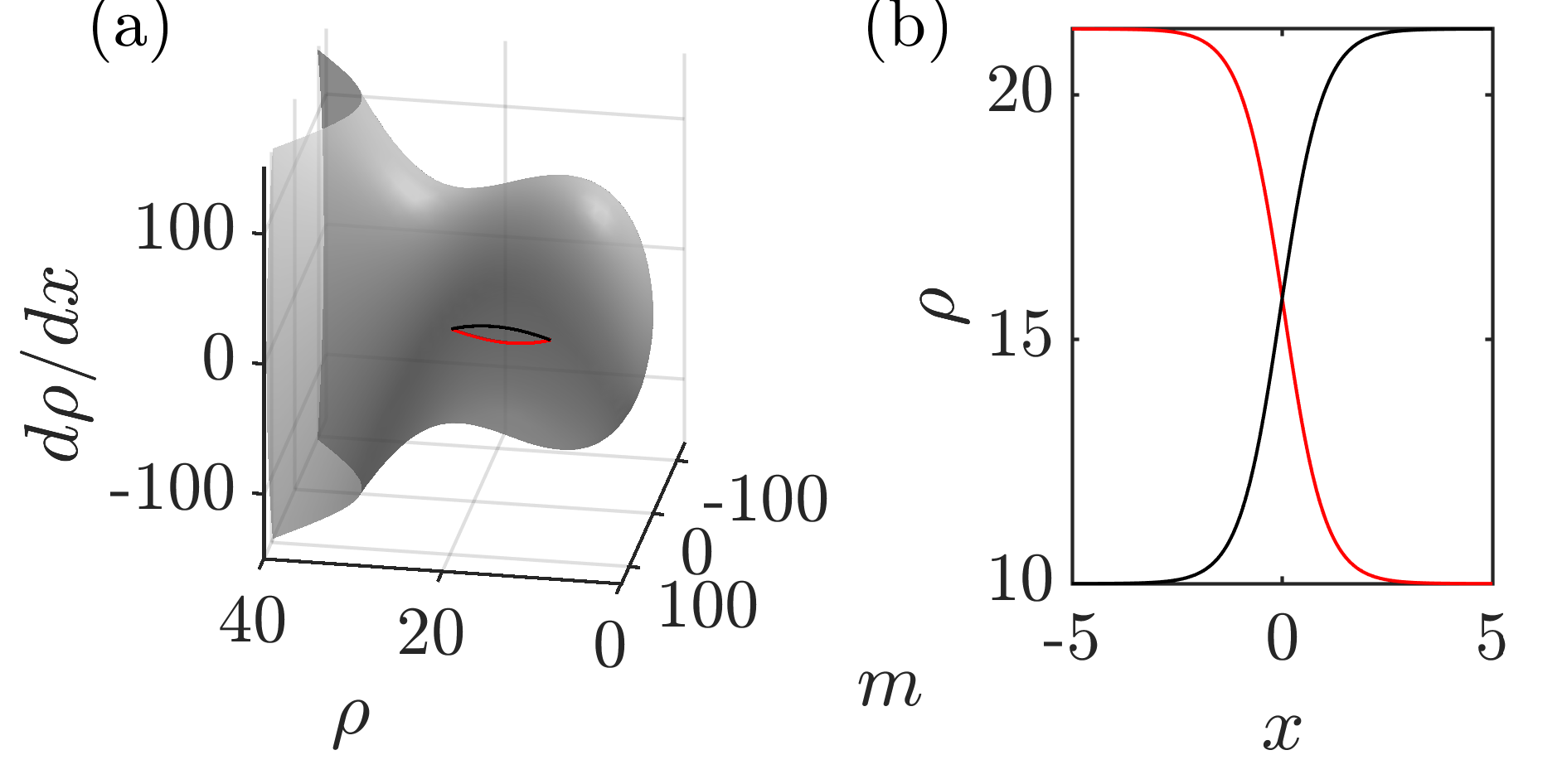}
    \includegraphics[width=0.4\linewidth]{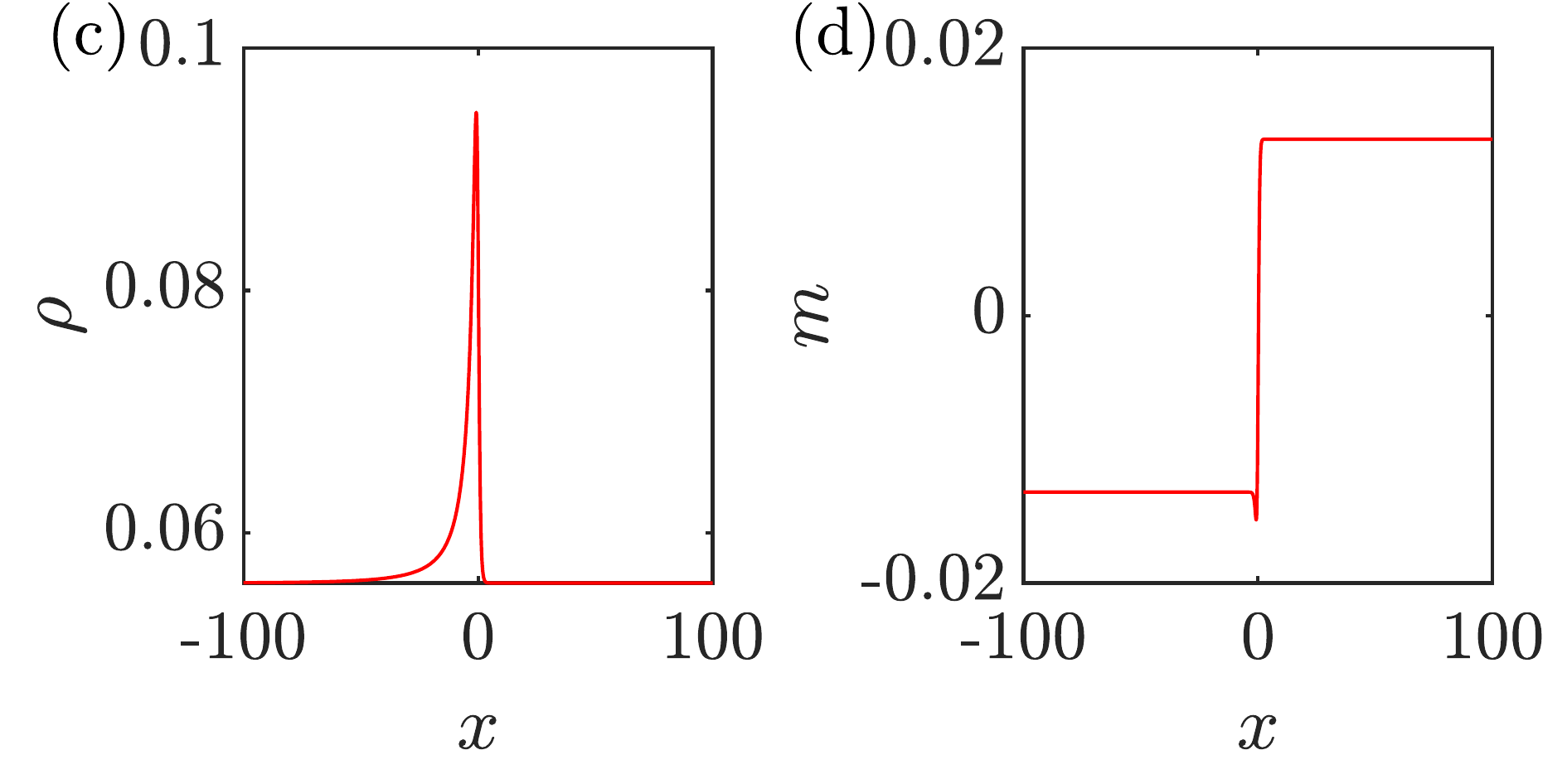}
    \caption{Kinks and antikinks for $w(x) = {\rm sech}^2(x)$.  (a) A
      type-II antikink (red solid) and kink (black solid) shown on the
      zero energy surface. (b) Corresponding density profiles. (c,d) A
      Type-I heteroclinic solution near where Type-II solutions emerge
      for small density.}
    \label{fig:limiting_kink}
\end{figure}

The end states of Type-II heteroclinic orbits lie between the maximum
of the connected energy surface
$ m^{({+})}=\sqrt{F(\rho)-\frac{1}{4}(\rho')^2}$ when
$\rho_{+}>\rho_c$ and between the saddle of
$m^{({+})}=\sqrt{F(\rho)-\frac{1}{4}(\rho')^2}$ and
$m^{({-})}=-\sqrt{F(\rho)-\frac{1}{4}(\rho')^2}$ for $\rho_+<\rho_c$
(see Fig.~\ref{fig:heteroclinic-solutions}(a)). The density $\rho_c$
is a critical density that depends on $w(x)$.  In the small density
regime $\rho_+\ll 1$, the critical points of $m^{({+})}$ and
$m^{({-})}$) coincide. This occurs along the sonic curve
$u_{+}^2 \rightarrow \rho_+$. In this limiting case, Type-II kinks
bifurcate from a type-I kink whose end states are related by
$u_{\pm}^2=\rho_{\pm}$.  See Fig.~\ref{fig:limiting_kink}(c,d).

\subsection{Summary of heteroclinic solutions}
\label{sec:summ-heter-solut}

\begin{figure}
  \centering
  \includegraphics[width=0.7\linewidth]{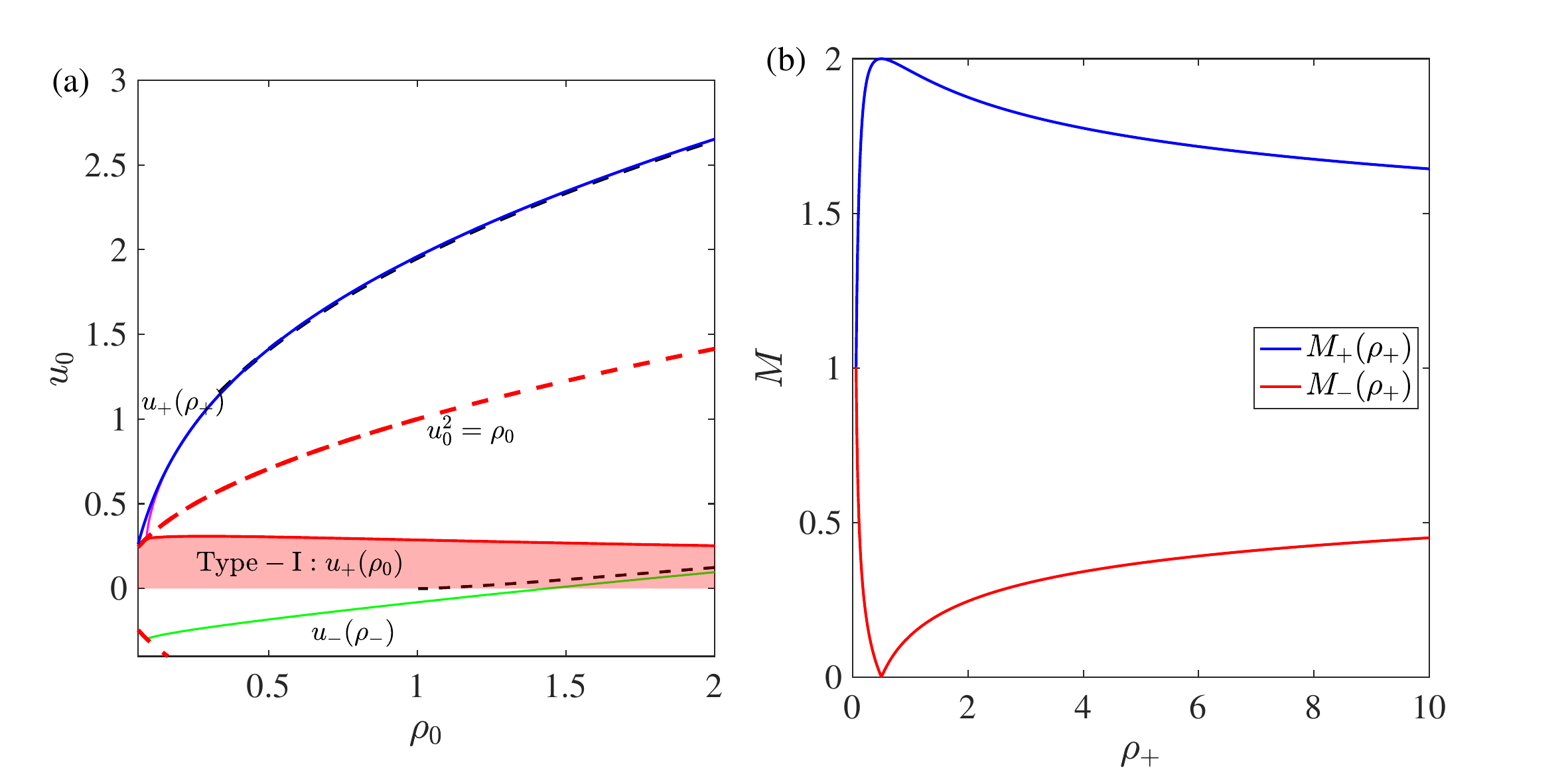}
  \caption{{Heteroclinic solutions with $w(x) = {\rm sech}(x)$ (a) and
    $w(x) = {\rm sech}^2(x)$ (b).  {(a)} Existence diagram for kink
    solutions when $w(x) = {\rm sech}(x)$. Existence region
    $u_+(\rho_0)$ for Type-I heteroclinics (red) and existence curves
    $u_-(\rho_-)$ (solid green) and $u_+(\rho_+)$ (solid magenta) for
    Type-II heteroclinics. Hydraulic curves $u_+ = \sqrt{f_2(\rho_+)}$
    and $u_- = \sqrt{f_3(\rho_-)}$ (dashed black) for Type-II
    heteroclinics.  The sonic curve (dashed red) is also
    shown. {(b)} The far-field Mach numbers
    $M_\pm = |u_\pm|/\sqrt{\rho_\pm}$ as a function of the
    continuation parameter $\rho_+$ for Type-II antikinks.}}
    \label{fig:heteroclinic_sech}
\end{figure}

% \begin{figure}
%     \centering
%     \includegraphics[width=0.5\linewidth]{figures/Fig_Mach_number.eps}
%     \caption{The Mach number distribution
%       $M_\pm = |u_\pm|/\sqrt{\rho_\pm}$ as a function of the
%       continuation parameter $\rho_+$ for Type-II antikinks with
%       $w(x)={\rm sech}^2(x)$.}
%     \label{fig:mach_number}
% \end{figure}

Figure \ref{fig:heteroclinic_sech}(a) is the existence diagram for
Type-I and Type-II antikinks when $w(x) = {\rm sech}(x)$.  The dashed
black curves show that hydraulic theory provides approximate existence
curves for a subset of Type-II kinks and antikinks but does not
account for Type-I kinks and antikinks.  Additionally, Type-II
antikinks extend to much smaller density than hydraulic theory would
suggest.  The existence curves in
Fig.~\ref{fig:heteroclinic-solutions} for $w(x) = {\rm sech}^2(x)$ are
qualitatively similar to those in Fig.~\ref{fig:heteroclinic_sech}.

The two-parameter family of Type-I kinks exhibit the same density and
speed but opposite velocities in the far-field, either flowing into or
out of the gain/loss region.  All Type-I kinks are found to be
subsonic.  On the other hand, the one-parameter family of Type-II
kinks and antikinks exhibits different densities and velocities in the
far-field with one of them being supersonic and the other subsonic as
in Fig.~\ref{fig:heteroclinic_sech}(b).

\section{Discussion and Conclusions}
\label{sec:discussion}

We have examined the existence and properties of stationary homoclinic
and heteroclinic solutions to the PT-symmetric defocusing NLS equation
\eqref{eq:1} with a complex potential by use of approximate and
numerical methods.  These solutions represent smooth transitions and
the scattering between two distinct plane waves
$\psi_\pm(x,t) = \sqrt{\rho_{\pm}} e^{i (u_\pm x - \mu t)}$ as
$x \to \pm \infty$ that are supported by the competition between
dispersion, nonlinearity, the repulsive potential
$\mathrm{Re}(V(x)) = -w(x)^2/2$, and gain/loss
$\mathrm{Im}(V(x)) = w'(x)/2$, reminiscent of dissipative solitons
\cite{akhmediev_dissipative_2008}.  The hydrodynamic interpretation in
which $\rho = |\psi|^2$ and {$u = \mathrm{Im} \,(\psi_x/\psi)$} are density and
velocity fields, respectively, provides insight into the nature of the
obtained solutions.  The stationary solutions to this system satisfy a
third-order, non-autonomous ODE with a pointwise conserved zero energy
equation that admits a variational formulation. The topology of the
associated energy surface depends on the far-field Mach number
$M = |u|/\sqrt{\rho}$ and can be connected or disconnected depending
upon whether $M < 2$ or $M > 2$ in the far-field, respectively.

We obtain a two-parameter family of homoclinic solutions, termed
depression waves, consisting of a localized density depletion in which
the subsonic far-field velocity accelerates in
the depleted region.  On the other hand, the
obtained two-parameter family of elevation waves exhibit a localized
density accumulation and remain subsonic throughout.  When the
far-field flow is sufficiently supersonic, accumulations in density
are found to resonate with stationary oscillations that persist in the
far-field, resulting in a three-parameter family of
generalized elevation waves on an oscillatory background.

Two classes of heteroclinic solutions are obtained with either the
same (Type-I) or different (Type-II) far-field densities and speeds.
In the former case, the far-field flows are equal and opposite and the
solution remains subsonic throughout.  We obtain a two-parameter
family of these Type-I heteroclinics.  The one-parameter family of
Type-II heteroclinic solutions represent transitions between sub and
supersonic far-field flows.  Type-II heteroclinics are found to result
from a symmetry breaking bifurcation of Type-I heteroclinics in the
small density regime.

These stationary solutions represent a significant generalization of
the family of constant intensity waves \cite{makris2015constant} and
can provide insight into initial value problems.  For example, in
\cite{chandramouli2023dispersive}, Type-I and Type-II heteroclinic
solutions of a very specific polarity were observed to emerge
dynamically from initial conditions to eq.~\eqref{eq:1} consisting of
constant intensity waves. This motivates further study of the
dynamical stability of these solutions.  In contrast to Type-II
solutions, Type-I heteroclinics are unique to the non-conservative
complex potential and have not been observed in the conservative
setting \cite{hakim_nonlinear_1997-1,leszczyszyn_transcritical_2009}.

Directions for future study include further dynamical systems analysis
to reveal the nature of bifurcations and possible additional solution
families.  The dynamical stability analysis of the obtained solutions
is a natural consideration, particularly because of the possibility of
the experimental realization of these states given recent advances in
non-Hermitian photonics
\cite{hang2017parity,hang2017realization,hang2013pt}. The underlying
symmetries considered here involve a spatially balanced gain–loss
distribution, accompanied by a symmetric, repulsive optical barrier
or, equivalently, a symmetric refractive index profile.

Another avenue of investigation includes the thorough characterization
of initial value problems for eq.~\eqref{eq:1} and their connection to
the obtained solutions, hinted at from preliminary work on this in
\cite{chandramouli2023dispersive}.  Understanding these stationary
solutions is a necessary step to explain the resonant generation of
unsteady dispersive hydrodynamic phenomena emerging from transcritical
flows in non-conservative and non-Hermitian systems.  In the context
of polariton condensates
\cite{amo_collective_2009,kamchatnov2012quasi}, the notion of
transcriticality has been found to be absent. Instead, a highly
oscillatory stationary pattern is seen to form upstream with no
downstream propagating disturbances.  Ultimately, we foresee that
studying the scattering of plane waves from a non-Hermitian potential
in the transcritical flow problem will inform the behavior of dark
soliton dynamics over nonconservative potentials. Of particular
interest is a detailed understanding of the waves reflected from, and
transmitted through a complex potential.

A non-trivial extension of this work is to address the types of
stationary solutions that arise for negative $u_0$ velocity parameters
and the subsequent transcritical flow problem. This scenario is unique
to the non-Hermitian case, as the conservative problem possesses a
reflection symmetry \cite{leszczyszyn_transcritical_2009}, whereas in
the Wadati case, the underlying spatial-reversal symmetry is broken
due to the presence of the gain-loss distribution.

\begin{acknowledgments}
  We thank Willy Hereman and Roberto Camassa
  for insightful discussions.  The authors would like to thank the Isaac Newton Institute for Mathematical Sciences, Cambridge, for support and hospitality during the programme Emergent phenomena in nonlinear dispersive waves, where work on this paper was undertaken. This work was supported by EPSRC grant EP/V521929/1.  MAH was partially supported by NSF grant DMS-2306319.
\end{acknowledgments}
%%%%%%%%%%%%%%%%%%%%%%%%%%%%%%%%%%%%%%%%%%%%%%%%%%%%%%%%%%%%%%%%%%%%%%

\appendix

\section{Analysis of the hydraulic approximation}
\label{sec:analys-hydr-appr}
Within this appendix, we present a more detailed discussion of the
roots of the quartic polynomial eq.~\eqref{eq:22} governing the
hydraulic regime.

\begin{figure}
  \centering
  \includegraphics[width=0.4\linewidth]{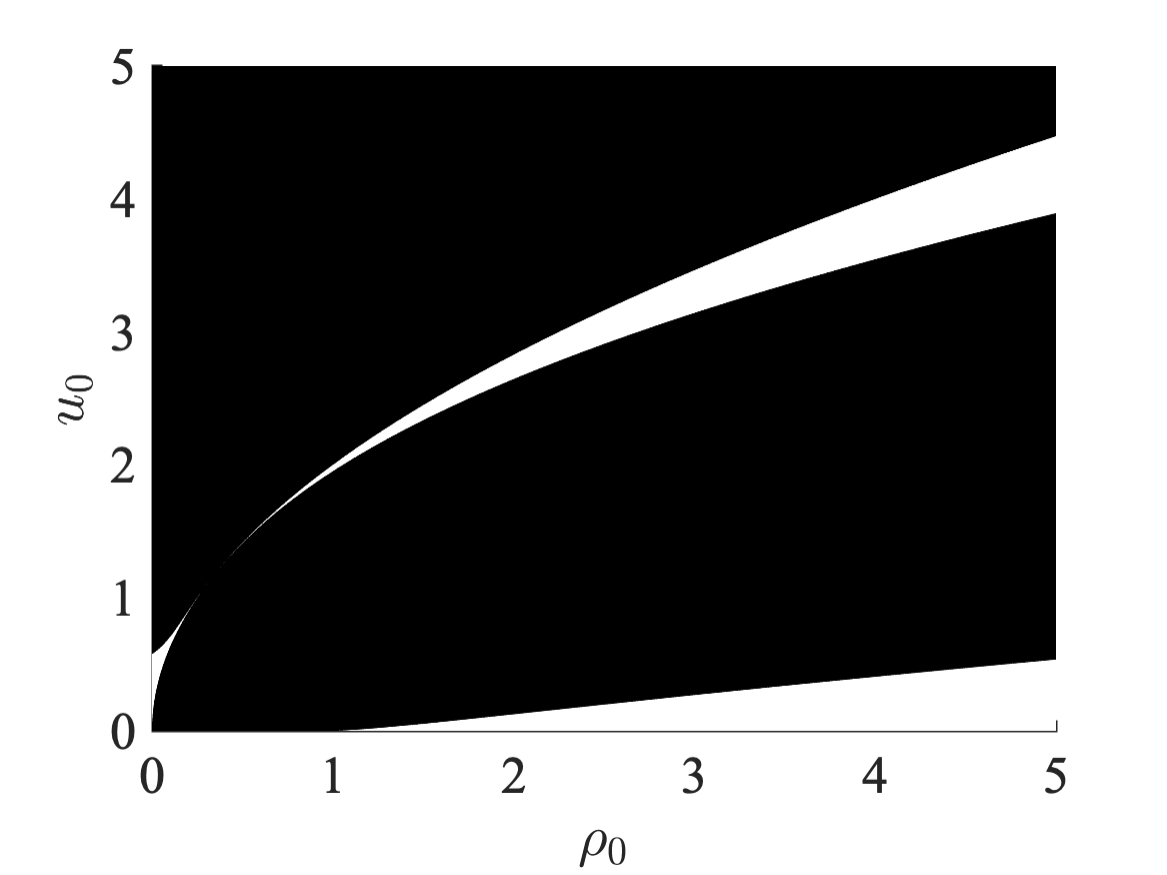}
  \caption{A plot showing the sign of the discriminant,
    ${\rm sgn}(D(0))$, of the quartic hydraulic equation
    [eq.~\eqref{eq:22}]. White (black) regions correspond to    ${\rm sgn}(D(0)) = 1$ (${\rm sgn}(D(0)) = -1$).}
    \label{fig:zero_discriminant}
\end{figure}
An examination of the roots of \eqref{eq:22} provides information on
existence and bifurcations of homoclinic solutions. Equation
\eqref{eq:22} has four roots we label
$\rho^{(a,b,c,d)} = \rho^{(a,b,c,d)}(X)$. The reality of these roots
depends on the discriminant $D$ of \eqref{eq:22}
\begin{equation}
  \label{eq:69}
  \begin{split}
    D(X) &= 4096 \rho_0^2 u_0^2 w(X)^4 \left ( 4\rho_0 - u_0^2 \right )
        \left ( \rho_0 + 2u_0^2 \right )^2K(X), \\
    K(X) &= 3(\rho_0 - u_0^2)^4 -4(2\rho_0 + u_0^2)(\rho_0^2 + 10 
        \rho_0 u_0^2 - 2 u_0^4)w(X)^2 + 6(\rho_0 - u_0^2)^2 w(X)^4 - w(X)^8 . 
  \end{split}
\end{equation}
If $D < 0$, two roots are real and the other two roots are complex
conjugates. When $D > 0$, the four roots are either all real or all
complex. Bifurcations in the number of roots can occur when $D = 0$.
The discriminant is zero for all $X$ when
\begin{equation}
  \label{eq:61}
  u_0^2 = f_1(\rho_0) \equiv 4\rho_0, \quad \mu^2=2A ,
\end{equation}
i.e., when $M_0 = |u_0|/\sqrt{\rho_0} = 2$.  This occurs precisely
when the zero energy surface's topology changes (see
Fig.~\ref{fig:zero_energy_surface}).  All other sign changes in $D$
occur in the factor $K(X)$.
% \begin{align}
%   \label{eq:spatial_discrim}
%   K &=  3(\rho_0 - u_0^2)^4 - 4(2\rho_0 + u_0^2)(\rho_0^2 + 10
%       \rho_0 u_0^2 - 2 u_0^4)w^2+ 6(\rho_0 - u_0^2)^2 w^4 - w^8,
% \end{align}
% Evaluating it at $X = 0$ gives
% \begin{align}
%   \label{eq:spatial_discrim_peak}
%   &K \big |_{w = 1} = 3(\rho_0 - u_0^2)^4 - 4(2\rho_0 + u_0^2)(\rho_0^2 + 10
%   \rho_0 u_0^2 - 2 u_0^4)+ 6(\rho_0 - u_0^2)^2 - 1,
      % \end{align}
We first observe that $K(X)$ exhibits a {local} minimum at $X = 0$ if and only
if
\begin{equation}
  \label{eq:27}
  (1-\rho_0)^2 - (1-10\rho_0)u_0^2 - 2 u_0^4 < 0 \quad \iff \quad
  u_0^2 < g(\rho_0) \equiv \frac{1}{4}\left ( 10\rho_0 - 1 + 3\sqrt{1
      - 4\rho_0+12\rho_0^2} \right ).
  % F(\rho_0,u_0)=4(2\rho_0 + u_0^2)(\rho_0^2 + 10 \rho_0 u_0^2 - 2
  % u_0^4)-12(\rho_0-u_0^2)^2+4\leq 0 .
\end{equation}
%\scn{\textbf{Note to be removed later}: $K'(X)=w'(X)\left(-8(2\rho_0+u_0^2)(\rho_0^2+10\rho_0 u_0^2-2u_0^4)w+24(\rho_0-u_0^2)^2w^3-8w^7\right)=0$ either at the root of $w'(X)=0$ or potentially at roots of the other polynomial in $w$. By Descarte's rule of signs, this polynomial (for $u_0<\sqrt{\frac{3\sqrt{3}}{2}+\frac{5}{2}}\sqrt{\rho_0}$ ) exhibits 2 sign changes and has 2 or 0 positive roots in $w$. This means there is a possibility of $K(X)$ having minima at other locations. Furthermore, $K(X)$ as a polynomial in $w$ exhibits 3 sign changes in $u_0<\sqrt{\frac{3\sqrt{3}}{2}+\frac{5}{2}}\sqrt{\rho_0}$, i.e. has either 3 or 1 positive real root(s). Also, we are interested in the roots within the interval $w\in[0,1]$.}
This is proven by evaluating $K'(0)$, $K''(0)$ and using $w(0) = 1$,
$w'(0)=0$ and $w''(0)<0$.  The expression $K(0)$ is itself a quartic
polynomial in $u_0^2$ whose discriminant is negative for
$\rho_0 \ge 0$, implying the existence of two real roots of $K(0)$
that we label
% Here, $F=0$ possesses a root
% $u_0^2=g(\rho_0)$, where $g(\rho_0)\geq f_1(\rho_0)$. Two of the roots
% of the quartic eq.~\eqref{eq:22} will be complex-valued when
% $u_0^2\geq g(\rho_0)$. However, when $u_0^2\leq g(\rho_0)$, $E(X)$ is
% minimized at $X=0$, which could indicate a change in the sign of
% $D$. The quantity $E|_{w=1}$ changes sign at its two roots given by
\begin{equation}
  \label{eq:59}
  u_0^2 \in \left \{ f_2(\rho_0),f_3(\rho_0) \right \} \quad
  \Rightarrow \quad K(0) = 0 \quad \Rightarrow \quad D(0) = 0 .  
\end{equation}
Both roots satisfy $f_j(\rho_0) \le g(\rho_0)$ so that $K(X)$ is
minimized {locally} at $X = 0$ for $u_0^2$ in the vicinity of and below these
roots.  By Descartes' rule of signs, only one of these roots
($f_2(\rho_0)$) is positive when $0 < \rho_0 < 1$; both roots are
positive when $\rho_0 > 1$.  The root $f_3(\rho_0) \lessgtr 0$ for
$\rho_0 \lessgtr 1$ can be approximated by
\begin{align}
  \label{eq:52}
  &f_3(\rho_0) \sim \left ( \frac{2\sqrt{6}}{9\rho_0^{1/4}}(\sqrt{\rho_0} -
    1)^{3/2} + \frac{11}{54 \sqrt{6} 
      \rho_0^{5/4}}(\sqrt{\rho_0}-1)^{7/2} \right )^2, \qquad   \rho_0  \to 1.
\end{align}
We also find that $f_1(\rho_0) \lessgtr f_2(\rho_0)$ for
$\rho_0 \lessgtr 1/3$ and
$f_1(1/3) = f_2(1/3) = g(1/3) = \frac{4}{3}$.

{The curves $u_0 \in \{\sqrt{f_j(\rho_0)}\}_{j=1}^3$ demarcating where
  the discriminant $D(0) = 0$ are shown in the bifurcation diagram of
  Fig.~\ref{fig:discriminant}, while Fig.~\ref{fig:zero_discriminant}
  is a contour map depicting the signs of $D(0)$ in the $\rho_0$-$u_0$
  plane. At first we note that, for the quartic polynomial
  \eqref{eq:22} when $|X| \to \infty$, $w(X) \to 0$ so we are
  interested in zeros of $\dot{F}(\rho)^2$. The discriminant of
  $\dot{F}(\rho)$ is $16\mu^2-24A=4(\rho_{0}-u_{0}^2)^2\geq 0$, and
  thus by extension, we assert the existence of real solutions in
  density to $\dot{F}(\rho)^2 = 0$ at $\pm \infty$. The corresponding
  velocity states are then computed using
  $u_{\infty}^2=F(\rho_{\infty})/\rho_{\infty}^2$. For each fixed
  $\mu$ and $A$, provided $\mu^2<2A$, it can be shown that there exist
  the following possible far-field solutions
  $u_{\infty}^2=\frac{2}{3}\left(-\mu \pm \sqrt{4\mu^2 - 6A}\right)$
  and $\rho_\infty=-\frac{2}{3}\mu \mp \frac{1}{3}\sqrt{4\mu^2 -
    6A}$. Thus, both far-field subsonic
  ($u_{\infty}^2/\rho_{\infty}<1$) and supersonic
  ($u_{\infty}^2/\rho_{\infty}>1$) conditions may be permitted when
  both pairs of double roots exist. However, when $\mu^2>2A$, the
  number of far-field double roots decreases from two to one where
  they coincide with $\rho_0$ and $u_0$ respectively:
  $u_\infty^2=\frac{2}{3}\left(-\mu +\sqrt{4\mu^2 - 6A}\right)=u_0^2$,
  $\rho_\infty=-\frac{2}{3}\mu - \frac{1}{3}\sqrt{4\mu^2 -
    6A}=\rho_0$. The spurious double roots describing subsonic
  far-field equilibria drop out. Thus, an analysis of the permissible
  far-field roots does not yield a definitive conclusion regarding the
  existence of homoclinic solutions satisfying the requisite boundary
  conditions.}  In particular, it might be necessary to examine $D(0)$
in various regions of the $\rho_0$-$u_0$ plane. This needs to be
investigated across the following intervals in $\rho_0$:
$\rho_0\leq 1/3$, $1/3<\rho_0<1$, and $\rho_0\geq 1$.
\begin{itemize}
\item $\rho_0\leq\frac{1}{3}$: Within this interval, the curves
  $u_0=\sqrt{f_{1,2}(\rho_0)}$ ($\sqrt{f_1}<\sqrt{f_2}$) divide the
  phase plane into distinct regions across which the signs of ${}D(0)$
  switch (see Fig.~\ref{fig:zero_discriminant}). Additionally, when
  $\mu^2>2A$ ($u_0^2>f_1(\rho_0)$), it was shown that the number of
  far-field double roots drops from 2 to 1. Focusing on
  $0<u_0<\sqrt{\rho_0}$, we note that $D(0)<0$, and thus two of the
  roots are complex conjugates. Additionally, two of the real-valued
  roots correspond to a depression wave with the incorrect far-field
  boundary condition
  $\rho_{\infty}^{(1)}=\frac{2 u_0^2+\rho_0}{3}<\rho_{\infty}^{(2)}$,
  where $\rho_{\infty}^{(2)}=\rho_0$ and an elevation wave on the
  correct density background $\rho_{\infty}^{(2)}$. Invoking the
  shifted momentum equation (eq.~\eqref{eq:9b})
  $2wm=-3(\rho-\rho_{\infty}^{(2)})\left(\rho-\rho_{\infty}^{(1)}\right)$,
  we see that the elevation wave $\rho>\rho_{\infty}^{(1,2)}$
  satisfies the incorrect boundary conditions in the shifted momentum
  ($m<0$). Across the curve $u_0=\sqrt{\rho_0}$, the roots of
  $\dot{F}(\rho)$ switch magnitudes:
  $\rho_{\infty}^{(1)}>\rho_{\infty}^{(2)}$. Now, a depression wave
  rests on the correct far-field density background
  $\rho_{\infty}^{(2)}$, with the shifted momentum profile being
  negative. At $u_0=2\sqrt{\rho_0}$ where $D(X)\equiv 0$, an elevation
  and depression root emerge at the background density $\rho_0$, with
  the elevation wave satisfying the correct boundary conditions in the
  shifted momentum. Two other double roots are obtained at
  $\rho_{\infty}^{(1)}$. The elevation wave satisfying the correct
  boundary conditions persists for $u_0>2\sqrt{\rho_0}$, while the two
  roots on the background $\rho_{\infty}^{(1)}$ become complex.
\item $\rho_0\in(1/3,1)$: The bifurcations within this density
  interval are similar to when $\rho_0\leq 1/3$, with an important
  difference arising from the ordering of the velocity curves
  $\sqrt{f_1}>\sqrt{f_2}$. Here, the elevation wave satisfying the
  correct boundary conditions emerges first across $\sqrt{f_2}$,
  together with an elevation-depression pair on the background
  $\rho_{\infty}^{(1)}$. The elevation root satisfying the correct
  boundary conditions is also seen to persist across
  $\sqrt{f_1}>\sqrt{f_2}$, even though two of the density roots on the
  background $\rho_{\infty}^{(1)}$ now become complex-valued.
\item $\rho_0> 1$. The bifurcation structure changes significantly
  within this density interval.  Here, $D(0)>0$ for
  $u_0<\sqrt{f_3(\rho_0)}$. All 4 roots are now real, with the
  depression wave on the $\rho_0$-background satisfying the correct
  shifted momentum boundary conditions (eq.~\eqref{eq:9b}). Across
  $\sqrt{f_3}$ two of the roots turn complex, one of them
  corresponding to the root satisfying the correct boundary
  conditions. A depression wave on $\rho_{\infty}^{(1)}$ and elevation
  wave on $\rho_{\infty}^{(2)}$ persist. {Moreover, at the bifurcation
    curve defined by $u_0=\sqrt{f_3(\rho_0)}$, the corresponding
    homoclinic solution loses smoothness at the peak of the
    inhomogeneity, an indicator of its eventual breakdown.}
  Eventually, across $\sqrt{f_2}$, $D(0)>0$, and the elevation and
  depression waves on $\rho_2=\rho_0$ emerge, satisfying the correct
  boundary conditions. Between $\sqrt{f_2}<u_0<\sqrt{f_1}$, all 4
  roots persist. Moreover, the elevation wave on $\rho_{\infty}^{(2)}$
  satisfying the correct boundary conditions and the depression on
  $\rho_{\infty}^{(2)}$ persist across $\sqrt{f_1}$, despite $D(0)<0$
  in this regime (see Fig.~\ref{fig:zero_discriminant}).
\end{itemize}
By stitching together the existence regimes for homoclinic solutions
satisfying the appropriate boundary conditions, we obtain simple
criteria summarizing their existence: either $D(0)>0$ for $\mu^2<2A$
($u_0<\sqrt{f_1}$) or $\mu^2>2A$ ($u_0>\sqrt{f_1}$).

We now study the roots $\rho^{(a,b,c,d)}(X)$ of \eqref{eq:22} near the
lowest bifurcation curve $u_0=\sqrt{f_3(\rho_0)}$ for
$\rho_0>1$. For $u_0 < \sqrt{f_3(\rho_0)}$, eq.~\eqref{eq:22}
possesses four rank-ordered distinct roots:
$\rho^{(a)} \leq \rho^{(b)}\leq \rho^{(c)} \leq \rho^{(d)}$. However,
only $\rho^{(c)}$ (see Fig.~\ref{fig:discriminant}(a)) respects the
boundary conditions $\rho^{(c)}(X) \to \rho_0$ and
$m^{(c)}(X)\to \rho_0 u_0$, as $X \to \pm \infty$ and corresponds to a
depression homoclinic in density. As
$u_0 \nearrow \sqrt{f_3(\rho_0)}$, the peaks of $\rho^{(c)}$ and
$\rho^{(b)}$ merge (see Fig.~\ref{fig:discriminant}(b)), and
kink-antikink profiles emerge.
% Importantly, the collision of the two
% roots is indicative of a saddle-node bifurcation in the underlying
% spatial dynamics, also observed in the conservative case
% \cite{leszczyszyn_transcritical_2009}.

Proceeding to the region bounded below by the curve
$u_0 = \sqrt{f_3(\rho_0)}$ and above by the curve
$u_0 = \sqrt{f_{1,2}(\rho_0)}$ where $D(0) < 0$
(cf.~Fig.\ref{fig:zero_discriminant}), only two roots of the quartic
\eqref{eq:22} are real and neither satisfies the correct boundary
conditions in density and shifted momentum.

At $u_0=\sqrt{f_2}$, $D(X)\geq 0$, four real roots re-emerge with two
merged into an antikink-kink-type profile
(Fig.~\ref{fig:discriminant}(c)). Crucially, of the roots merged into
a kink-antikink pair, the elevation wave satisfies the correct density
and shifted momentum boundary conditions. A further increase in $u_0$
results in the persistence of the distinct elevation real root
$\rho^{(b)}(X)$ that satisfies the boundary conditions. This is shown
in Fig.~\ref{fig:discriminant}(d).  Approximate formulas for the
homoclinic orbits can be found in the far-field where $w(X)\ll
|\mu|$. Expanding the density and the momentum in a power series in
$w(X)$, the root satisfying the boundary conditions is
\begin{align}
  \label{eq:46}
  \rho(X) &= \rho_0 - w(X) \frac{\rho_0 u_0}{\rho_0 - u_0^2} -
            w^2(X)\frac{3u_0^2\rho_0^2}{2(\rho_0-u_0^2)^3} + \cdots, 
  \\\nonumber m(X) &= \rho_0 u_0 + w(X)^2 \frac{\rho_0 u_0}{2(\rho_0 - u_0^2)} + 
                     \cdots,
\end{align}
corresponding to an elevation (depression) homoclinic when
$u_0 \gg \sqrt{\rho_0}$ ($u_0\ll\sqrt{\rho_0}$).

For the case when $u_0 = \sqrt{f_1}(\rho_0) = 2 \sqrt{\rho_0}$ and
$\mu^2 = 2 A$, the root can be written compactly as
\begin{equation}
  \label{eq:hydraulic_elevation_explicit}
  \rho(X)=\frac{1}{9}\left(\sqrt{12 \sqrt{2A} w^2(X) + 4w^4(X)} +
    3\sqrt{2A} + 2w^2(X)\right),\;\;m(X)=\sqrt{F(\rho(X))}.
\end{equation}

\section{Hydraulic analysis of the conservative case with repulsive potential}
\label{Hydraulic-homoclinic-repulsive}

We now consider the conservative NLS equation \eqref{eq:1} in which
$V(x)=\sigma w^2(x)/2$, $\sigma=\pm 1$. The case when $\sigma=1$ has
been explored extensively and characterized in
\cite{hakim_nonlinear_1997,leszczyszyn_transcritical_2009}. We study
the case when $\sigma=-1$, which aligns with the sign of the real part
of the Wadati potential \eqref{eq:58}.

The search for stationary solutions via a hydraulic analysis in which
$X=\epsilon x$ results in the system of equations for homoclinic
solutions
\begin{align}
  \rho u= \rho_0 u_0,\;\;
  \frac{1}{2}u^2+\rho-\frac{1}{2}w^2=\frac{1}{2}u_0^2+\rho_0,
\end{align}
where $u(X)\rightarrow u_0$ (and $\rho(X)\rightarrow \rho_0$) as
$|X|\rightarrow \infty$.  Combining the two equations, we obtain a
depressed cubic equation in $u(X)$ given by
\begin{equation}
\label{Cubic-eqn}
   u^3-(u_0^2+2\rho_0+w^2)u+2\rho_0 u_0=0.
\end{equation}
To assess the nature of the roots, we study the discriminant of the
cubic equation, given by
\begin{equation}
    D(X)=4\left(u_0^2+2\rho_0+w^2\right)^3-108\rho_0^2u_0^2.
\end{equation}
In particular, given the positive definiteness of $w(X)>0$, it can be
shown that the spatial discriminant is always positive when
$\rho_0>0$. Thus, all three roots of the cubic equation in
eq.~\eqref{Cubic-eqn} are real-valued and do not coalesce for any
finite $X$, distinguishing this case from the conservative case with
attractive potential. This analysis reveals the crucial role played by
the imaginary part of the potential \eqref{eq:58} in eq.~\eqref{eq:1}
when it comes to determining the existence of homoclinic solutions.

% The transcritical flow problem has been studied for the
% Gross-Pitaevskii equation \eqref{eq:1} with real-valued attractive
% potential $V$ in
% \cite{hakim_nonlinear_1997,leszczyszyn_transcritical_2009}.  These
% results helped explain the dynamics observed in BEC
% \cite{engels_stationary_2007} and photorefractive optical media
% \cite{wan_wave_2010} experiments.  In this case, the boundary
% conditions consist of a nonlinear plane wave solution of
% eq.~\eqref{eq:1} whose wavevector corresponds to the flow velocity.
% The simulations in \cite{chandramouli2023dispersive} motivate a
% nonconservative generalization of the transcritical flow problem to
% non-Hermitian media. In the absence of the gain-loss term, a repulsive
% potential permits the existence of homoclinic flow patterns for
% arbitrary freestream boundary conditions
% \cite{leszczyszyn_transcritical_2009}.
% % (see App. \ref{Hydraulic-homoclinic-repulsive} for an analysis using
% % \textit{hydraulic} theory).
% In contrast, the conservative case with attractive potential does not
% appear to support 
% \cite{hakim_nonlinear_1997,leszczyszyn_transcritical_2009}.  This
% (non)existence of homoclinic flow patterns has important implications
% for the resonant generation of DSW.

\section{Numerical methods}
\label{sec:numerical-methods}

\textbf{Depression waves:} The computation of depression waves is
achieved utilizing a pseudospectral-Fourier discretization.  We use
two distinct solution approaches in order to verify existence.  In the
first, we apply nonlinear least-square minimization of the residual at
equispaced points using MATLAB's \texttt{fsolve} (Levenberg-Marquardt
solver) routine to the first of eq.~\eqref{eq:29} with
$m(x) = \sqrt{F(\rho(x))-\frac{1}{4}\rho'(x)^2}$ from the zero energy
equation \eqref{eq:62}.  The domain length is $L = 20$ and $2^8$
  points are used. The residual for these solutions is below
$10^{-8}$. We then numerically verify that the parametric equation
$\dot m(\rho)=-w$ is satisfied to the same tolerance.  A second,
independent means of computing solutions applies the Newton
bi-conjugate gradient algorithm
\cite{yang_newton-conjugate-gradient_2009} to the full third order
system in eq.~\eqref{eq:29} with the residual below $10^{-8}$.  

We compute the $u_0$ intercept $\rho_*$ by numerically computing even
solutions to eq.~\eqref{Scalar-ODE-Bifurcation-from-CI-limit}
utilizing Matlab's \texttt{bvp5c} subject to the boundary conditions
$u_1(-L) = 0$, $u_1'(0)=0$ and $u_1(0)=1$. Thereafter, we check that
the density is localized by calculating $\rho_1(-L)$.  Spanning the
interval in $\rho_0\in [0,1]$, we utilize a parametric discretization
of $\Delta \rho_0=10^{-4}$.  For each $n$, there is a distinct
$\rho_* = \rho_0$ for which the correct far-field behavior for $\rho$
is no longer satisfied.

\textbf{Elevation waves:} We compute elevation waves in the small
density regime by recasting the ODE \eqref{eq:29} as the
integro-differential equation
\begin{equation}
  \label{Volterra-type-for-elevation-homoclinics}
  \rho''(x)-2\dot{F}(\rho)-4w(x)\left(\rho_0
    u_0+\int_{-\infty}^{x}(-w(y) \rho'(y)) dy\right)=0 .
\end{equation}
We use a pseudospectral-Fourier discretization and Matlab's
\texttt{fsolve} applied to
eq.~\eqref{Volterra-type-for-elevation-homoclinics} evaluated at
equispaced points.  We perform continuation in $u_0$ from the CI wave
solution \eqref{eq:12} for fixed $0 < \rho_0 < \rho_*$.  The integral
term in eq.~\eqref{Volterra-type-for-elevation-homoclinics} is
computed with the FFT; see \cite{el_dispersive_2017-1}.

\textbf{Generalized elevation waves:} Assuming even parity, these
solutions satisfy the boundary condition \eqref{eq:88} and
$m(x) \to \rho_0/\varepsilon$ as $|x| \to \infty$. For a fixed
$\sigma \in (0,\pi/2]$, we use the computational domain length
$2L=2(N+\sigma/\pi)\pi/k$ and determine the unknown amplitude $R_0$ at
$\pm L$ using a shooting procedure \cite{yang_nonlinear_2010} on the
ODE system Eq.~\eqref{eq:29} that enforces the symmetry condition
$\zeta'(0)=0$.  Note that $k_{\varepsilon} = 2/\varepsilon = 2|u_0|$
is the approximate wavenumber of the resonant background. To obtain
the numerical solutions, we use the exact resonant wavenumber
$k = 2\sqrt{u_0^2-\rho_0}$ for the computations. Here, $N$ is a large,
positive integer. The Matlab routine \texttt{fsolve} is used to obtain
$R_0$ to high accuracy with residual $\sim 10^{-15}$.  Once the tail
amplitude is obtained, the generalized elevation wave is obtained by
solving the ODEs \eqref{eq:29} with initial conditions
$\rho(-L) = \rho_0 + R_0 \cos(k x+\sigma)$ and
$m(-L)=\rho_0 u_0$.

\textbf{Type-I kinks:} We numerically solve the integro-differential
equation
\begin{equation}
  \label{Volterra-type-for-elevation-TypeI}
  \rho''(x)-2\dot{F}(\rho)-4w(x)\left(\rho_{-}
    u_{-}+\int_{-\infty}^{x}(-w(y)
    \rho'(y)) \, \mathrm{d}y\right)=0 ,
\end{equation}
that is equivalent to eq.~\eqref{eq:29} subject to the heteroclinic
boundary conditions \eqref{eq-hetero-1} for $x \in [-L,L]$ with
$L = 200$.  We use a pseudospectral-Fourier discretization scheme to
eq.~\eqref{Volterra-type-for-elevation-TypeI} with $N = 2^{11}$
equispaced points and apply Matlab's \texttt{fsolve} to the resulting
nonlinear system of equations with residuals below {$10^{-9}$}.

We also compute a two-parameter family of Type-I antikinks using
numerical continuation in the parameters $u_+$ and $\rho_0$. First, a
density continuation from $\rho_0=0.25$ is performed upon fixing the
right background velocity $u_+=-u_{-}\approx 0.0519$ and computing
solutions in the density interval $\rho_0\in [0.06,3)$ in steps of
$\Delta \rho_0=0.01$. Thereafter, a continuation in $u_+$ in steps of
$\Delta u_+= 10^{-4}$ is performed. The velocity continuation
terminates at a turning point.

\textbf{Type-II kinks:} For the computation of antikinks, we formulate
a two-point boundary value problem for $\mathbf{q} = (m,\rho,\rho')^T$
by fixing the right density state to a given value $\rho_+$ together
with Neumann boundary conditions enforced at either end for the
density profile
\begin{align}
  \label{eq:33kink}
    &\mathbf{q}' =
    \begin{pmatrix}
      -w(x) q_3 \\ q_3 \\ 2 \dot{F}(q_2) + 4 w(x) q_1
    \end{pmatrix}, \quad x \in (-L,L), \quad
    \begin{pmatrix}
      q_2(L) \\ q_3(-L) \\ q_3(L)
    \end{pmatrix} =
    \begin{pmatrix}
      \rho_+ \\ 0 \\ 0
    \end{pmatrix} .
\end{align}
The right asymptotic state in the hydrodynamic velocity is solved
together with the solution profile, which is done by enforcing the
jump conditions \eqref{eq-hetero-2} in $A$ and $\mu$, to an absolute
tolerance of $10^{-12}$ in Matlab's \texttt{bvp5c}.  The initial guess
is the hydraulic solution for large $\rho_+$.  We then perform
continuation in $\rho_+$ to the small density regime. As a secondary
validation step, we verified that solutions lie on the zero energy
surface \eqref{eq:17}.

\bibliography{Wadati_file_MH}% Produces the bibliography via BibTeX.
%%%%%%%%%%%%%%%%%%%%%%%%%%%%%%%%%%%%%%%%%%%%%%%%%%%%%%%%%%%%%%%%%%%%%%%%%%%%%%%%%%%%
\end{document}